\documentclass{aa}
\usepackage{graphicx}
\usepackage{txfonts}
\usepackage{subfig}

\usepackage{longtable}
\usepackage{tabularx}
\usepackage{pdflscape}
\usepackage{multicol}

\usepackage{hyperref}
 \hypersetup{
     colorlinks=true,
     linkcolor=blue,
     filecolor=magenta,
     urlcolor=cyan,
     citecolor=cyan,
 }
\usepackage{epsf}
\usepackage{rotating}

\usepackage{xcolor}

\usepackage[multiple]{footmisc}

\newcommand{\footlabel}[2]{%
    \addtocounter{footnote}{1}%
    \footnotetext[\thefootnote]{%
        \addtocounter{footnote}{-1}%
        \refstepcounter{footnote}\label{#1}%
        #2%
    }%
    $^{\ref{#1}}$%
}
\newcommand{\teff}{\mbox{$T_{\rm eff}$}}
\newcommand{\logg}{\mbox{$\log g$}}
\newcommand{\vsini}{\mbox{$v \sin i_*$}}
\newcommand{\feh}{[Fe/H]}

\newcommand{\kms}{\mbox{km\,s$^{-1}$}}
\newcommand{\ms}{\mbox{m\,s$^{-1}$}}
\newcommand{\halpha}{\mbox{$H_\alpha$}}
\newcommand{\rv}{RV}
\newcommand{\ccf}{CCF}
\newcommand{\bis}{BIS}
\newcommand{\fwhm}{FWHM}
\newcommand{\snr}{S/N}
\newcommand{\logrhk}{\mbox{$\log\textrm{R'}_{\textrm{HK}}$}}
\newcommand{\sophie}{\textsc{sophie}}
\newcommand{\sophiep}{\textsc{sophie+}}
\newcommand{\elodie}{\textsc{elodie}}

\newcommand{\glsp}{GLSP}
\newcommand{\fap}{\textsc{fap}}
\newcommand{\mjup}{\mbox{$\mathrm{M}_{\rm jup}$}}
\newcommand{\mpl}{\mbox{$M_{\rm p}$}}

\newcommand{\dispratio}[1]{\mbox{$\frac{\mathrm{std}(\mathrm{ #1 })}{<\sigma_{\mathrm{#1}}>}$}}

\newcommand{\pdf}{\textsc{pdf}}
\newcommand{\prior}{prior}
\newcommand{\post}{posterior}

\renewcommand{\eqref}[1]{\ref{eq:#1}}

\newcommand{\sectionname}{Section}
\newcommand{\sectref}[1]{\ref{sect:#1}}
\newcommand{\Sect}[1]{\sectionname~\sectref{#1}}
\newcommand{\sect}[1]{\Sect{#1}}
\newcommand{\sectlabel}[1]{\label{sect:#1}}
\newcommand{\sectalt}[1]{\sectref{#1}}
\newcommand{\tabname}{Table}
\newcommand{\tabref}[1]{\ref{tab:#1}}
\newcommand{\Tab}[1]{\tabname~\tabref{#1}}
\newcommand{\tab}[1]{\Tab{#1}}
\newcommand{\tablabel}[1]{\label{tab:#1}}

\newcommand{\figname}{Fig}
\newcommand{\figref}[1]{\ref{fig:#1}}
\newcommand{\Fig}[1]{\figname~\figref{#1}}
\newcommand{\fig}[1]{\Fig{#1}}
\newcommand{\figlabel}[1]{\label{fig:#1}}
\newcommand{\figalt}[1]{\figref{#1}}
\newcommand{\appname}{Appendix}
\newcommand{\appref}[1]{\ref{app:#1}}
\newcommand{\App}[1]{\appname~\appref{#1}}
\newcommand{\app}[1]{\App{#1}}
\newcommand{\applabel}[1]{\label{app:#1}}

\begin{document}

\title{The SOPHIE search for northern extrasolar planets
}

\subtitle{XVIII. Six new cold Jupiters, including one of the most eccentric exoplanet orbits\thanks{Tables \tabref{rvs27969} to \tabref{rvs211403} 
are only available in electronic form at the CDS via anonymous ftp to cdsarc.u-strasbg.fr (130.79.128.5) or via \url{http://cdsweb.u-strasbg.fr/cgi-bin/qcat?J/A+A/}}
}
\titlerunning{Six new cold Jupiters}
\authorrunning{Demangeon et al.}

\author{Olivier D. S. Demangeon\inst{\ref{IA-Porto},\ref{UP}, \thanks{\email{olivier.demangeon@astro.up.pt}}}
\and S. Dalal\inst{\ref{IAP}}
\and G. H\'ebrard\inst{\ref{IAP},\ref{OHP}}
\and B. Nsamba\inst{\ref{MPIA-Garching},\ref{Uganda},\ref{IA-Porto}}
\and F. Kiefer\inst{\ref{IAP},\ref{LESIA}}
\and J. D. Camacho\inst{\ref{IA-Porto},\ref{UP}}
\and J. Sahlmann\inst{\ref{RHEA-ESAC}}
\and L. Arnold\inst{\ref{OHP}}
\and N. Astudillo-Defru\inst{\ref{DepMatFis-UCSC}}
\and X. Bonfils\inst{\ref{Uni-Grenoble}}
\and I. Boisse\inst{\ref{Uni-AixMarseille-LAM}}
\and F. Bouchy\inst{\ref{Geneve-obs}}
\and V. Bourrier\inst{\ref{Geneve-obs}}
\and T. Campante\inst{\ref{IA-Porto},\ref{UP}}
\and X. Delfosse\inst{\ref{Uni-Grenoble}}
\and M. Deleuil\inst{\ref{Uni-AixMarseille-LAM}}
\and R. F. D\'{\i}az\inst{\ref{ICAS}}
\and J. Faria\inst{\ref{IA-Porto},\ref{UP}}
\and T. Forveille\inst{\ref{Uni-Grenoble}}
\and N. Hara\inst{\ref{Geneve-obs}}
\and N. Heidari\inst{\ref{Uni-AixMarseille-LAM},\ref{Tehran},\ref{OCA}}
\and M. J. Hobson\inst{\ref{IA-UCSC},\ref{MIA}}
\and T. Lopez\inst{\ref{Uni-AixMarseille-LAM}}
\and C. Moutou\inst{\ref{Uni-Toulouse-IRAP}}
\and J. Rey\inst{\ref{LasCampanas}}
\and A. Santerne\inst{\ref{Uni-AixMarseille-LAM}}
\and S. Sousa\inst{\ref{IA-Porto}}
\and N. C. Santos\inst{\ref{IA-Porto},\ref{UP}}
\and P. A. Str\o m\inst{\ref{IAP},\ref{Warwick}}
\and M. Tsantaki\inst{\ref{INAF-Firenze},\ref{IA-Porto}}
\and S. Udry\inst{\ref{Geneve-obs}}
}

\institute{
% Portugal
Instituto de Astrof\'{\i}sica e Ci\^encias do Espa\c co, CAUP, Universidade do Porto, Rua das Estrelas, 4150-762, Porto, Portugal
\label{IA-Porto}
% \and Centro de Astrof\'{\i}sica da Universidade do Porto, Rua das Estrelas, 4150-762 Porto, Portugal
% \label{CAUP}
\and Departamento de F\'{\i}sica e Astronomia, Faculdade de Ci\^encias, Universidade do Porto, Rua do Campo Alegre, 4169-007, Porto, Portugal
\label{UP}
% France
\and Institut d'Astrophysique de Paris, 98 bis, boulevard Arago, 75014, Paris
\label{IAP}
\and Observatoire de Haute-Provence, CNRS, Universit\'e d'Aix-Marseille, 04870, Saint-Michel-l'Observatoire, France
\label{OHP}
\and LESIA, Observatoire de Paris, Université PSL, CNRS, Sorbonne Université, Université de Paris, 5 place Jules Janssen, 92195, Meudon, France
\label{LESIA}
\and Univ. Grenoble Alpes, CNRS, IPAG, 38000 Grenoble, France
\label{Uni-Grenoble}
\and Aix-Marseille Univ, CNRS, CNES, LAM, Marseille, France
\label{Uni-AixMarseille-LAM}
\and Univ. de Toulouse, CNRS, IRAP, 14 Avenue Belin, 31400, Toulouse, France
\label{Uni-Toulouse-IRAP}
\and Laboratoire J.-L. Lagrange, Observatoire de la C\^ote d’Azur (OCA), Universite de Nice-Sophia Antipolis (UNS), CNRS, Campus Valrose, 06108 Nice Cedex 2, France.
\label{OCA}
% Switzerland
\and Observatoire de Gen\`eve,  Universit\'e de Gen\`eve, Chemin Pegasi, 51, 1290 Sauverny, Switzerland
\label{Geneve-obs}
% Spain
\and RHEA Group for the European Space Agency (ESA), European Space Astronomy Centre (ESAC), Camino Bajo del Castillo s/n, 28692 Villanueva de la Ca\~nada, Madrid, Spain
\label{RHEA-ESAC}
% Chile
\and Departamento de Matem\'atica y F\'isica Aplicadas, Universidad Cat\'olica de la Sant\'isima Concepci\'on, Alonso de Rivera 2850, Concepci\'on, Chile
\label{DepMatFis-UCSC}
\and Instituto de Astrofésica, Pontificia Universidad Católica de Chile, Av. Vicuña Mackenna 4860, Macul, Santiago, Chile
\label{IA-UCSC}
\and Millennium Institute of Astrophysics, Av. Vicuña Mackenna 4860, 7820436 Macul, Santiago, Chile
\label{MIA}
\and Las Campanas Observatory, Carnegie Institution of Washington, Colina el Pino, Casilla 601 La Serena, Chile
\label{LasCampanas}
% Argentina
\and International Center for Advanced Studies (ICAS) and ICIFI (CONICET), ECyT-UNSAM, Campus Miguelete, 25 de Mayo y Francia, (1650) Buenos Aires, Argentina
\label{ICAS}
% UK
\and Department of Physics, University of Warwick, Coventry, CV4 7AL, UK
\label{Warwick}
% Germany
\and Max-Planck-Institut für Astrophysik, Karl-Schwarzschild-Str. 1, D-85748 Garching, Germany
\label{MPIA-Garching}
% Italy
\and INAF – Osservatorio Astrofisico di Arcetri, Largo E. Fermi 5, 50125 Firenze, Italy
\label{INAF-Firenze}
% Iran
\and Department of Physics, Shahid Beheshti University, Tehran, Iran.
\label{Tehran}
% Uganda
\and Mbarara University of Science and Technology, P. O Box 1410, Mbarara - Uganda
\label{Uganda}
}

\date{Received date / Accepted date }

\abstract
% % context heading (optional)
{Due to their low transit probability, the long-period planets are, as a population, only partially probed by transit surveys. Radial velocity surveys thus have a key role to play, in particular for giant planets. Cold Jupiters induce a typical radial velocity semi-amplitude of 10\,\ms, which is well within the reach of multiple instruments that have now been in operation for more than a decade.
}
%% aims heading (mandatory)
{We take advantage of the ongoing radial velocity survey with the \sophie\ high-resolution spectrograph, which continues the search started by its predecessor \elodie\ to further characterize the cold Jupiter population.
}
% % methods heading (mandatory)
{Analyzing the radial velocity data from six bright solar-like stars taken over a period of up to 15 years, we attempt the detection and confirmation of Keplerian signals.
}
% % results heading (mandatory)
{We announce the discovery of six planets, one per system, with minimum masses in the range 2.99-8.3\,\mjup\ and orbital periods between 200 days and 10 years. The data do not provide enough evidence to support the presence of additional planets in any of these systems. The analysis of stellar activity indicators confirms the planetary nature of the detected signals.
}
%% conclusions heading (optional), leave it empty if necessary
{These six planets belong to the cold and massive Jupiter population, and four of them populate its eccentric tail. In this respect, HD 80869 b stands out as having one of the most eccentric orbits, with an eccentricity of $0.862^{+0.028}_{-0.018}$. These planets can thus help to better constrain the migration and evolution processes at play in the gas giant population. Furthermore, recent works presenting the correlation between small planets and cold Jupiters indicate that these systems are good candidates to search for small inner planets.
}


\maketitle

\section{Introduction}\sectlabel{intro}

Transit photometry surveys, and in particular Kepler \citep{Borucki-2010}, have revolutionized our understanding of planetary systems. With the discovery of more than 4500 transiting planet candidates\footnote{The planet candidates discovered by the Kepler mission are available at the \href{https://exoplanetarchive.ipac.caltech.edu/cgi-bin/TblView/nph-tblView?app=ExoTbls&config=cumulative&constraint=koi_pdisposition+like+\%27CANDIDATE\%27}{NASA Exoplanet Archive}.}, Kepler offered a statistically complete view of the radius distribution of the exoplanet population out to orbital periods of about 100 days \citep{thompson2018}. However, due to the low transit probability and the limited duration of the survey (4 years), Kepler provides only a partial view of the exoplanet population with orbital periods between 100 days and 4 years \citep{foreman-mackey2016,hsu2019a} and is nearly blind for longer periods. Radial velocity (\rv) surveys thus play a pivotal role in assessing the properties of the exoplanet population in this period range \citep{dalba2021}.

Combining the detections provided by Kepler with the ones from the \textsc{harps} and \textsc{coralie} \rv\ surveys \citep{mayor2011}, \citet{fernandes2019} inferred the frequency of giant planets at long orbital periods\footnote{\citet{fernandes2019} defined cold Jupiters as planets with a mass or a minimum mass in the range of 0.1-20 \mjup\ and a semimajor axis between 0.1 and 100\,au.}, hereafter cold Jupiters, to be $26.6^{+7.5}_{-5.4}\,\%$. More precisely, their frequency rises out to the snow line, $\sim$ 2-3\,au ($\sim$1000-2000 days of orbital period), as previously reported by other \rv\ surveys \citep[][]{cumming2008} but then decreases as the orbital period increases. After the snow line, the frequency of giant planets is only a few percent \citep{wittenmyer2016}. Extrapolating these results to even larger separations, 10\, au and above, provides a good agreement with the findings of direct imaging surveys, which estimate the frequency of giant planets at large separation to be $0.6^{+0.7}_{-0.5}\,\%$ \citep{bowler2016}. \citet{fernandes2019} also compared the observed frequency distribution with several population synthesis models \citep[][]{ida2018,mordasini2018,jennings2018} and concluded that the core-accretion formation scenario coupled with Type-II migration produces the observed turnover around the snow line. Furthermore, models that simulate the formation of multiple giant planets and the interaction between them produce the best agreement with the observed distributions.

Besides providing valuable observational constraints on the formation of giant planets, the study of the cold Jupiter population is also essential for understanding the properties of the inner planetary systems. Giant planets are thought to be the first planets to form \citep[within the first 10\,Myr;][]{pascucci2006}, when gas is still present in the protoplanetary disk. The time at which terrestrial planets start to appear is still debated in the literature \citep[e.g.,][]{raymond2005, lammer2021}, but it is thought to be either contemporaneous to or later than the apparition of giant planets, which can thus be heavily influenced by their lighter siblings \citep[e.g.,][]{morbidelli2012,Levison2003}.

The \sophie\ spectrograph, mounted at the 1.93-meter telescope of the Observatoire de Haute-Provence (France), and its predecessor \elodie\ have been used for surveys dedicated to the search for exoplanets in \rv\ since this search began. \elodie\ was the instrument used for the discovery of the first exoplanet around a solar-type star in 1995 \citep{Mayor-1995}. We present here new results from the \rv\ survey of giant planets that we are conducting with the \sophie\ spectrograph in a volume-limited sample of solar-type stars \citep[e.g.,][]{moutou2014,diaz2016,hebrard2016}. Some of the targets presented in this work were already observed with \elodie, and here we provide time baselines of up to 15 years. Such data sets are gold mines for detecting new cold Jupiters and refining our understanding of this population. In \sect{rvdatasets} we present the \rv\ data sets. In \sect{star} and \sectref{planet} we explain our derivation of the host star and planet properties. Finally, in \sect{discussion} we put the newly detected cold Jupiters in perspective and emphasize their contribution to our understanding of planetary systems.

\section{The data sets: High-resolution spectra and radial velocities}\sectlabel{rvdatasets}

In this section, we present the high-resolution spectra of six stars in our sample obtained with \sophie\ and \elodie. \elodie\ has a resolution of 42000 \citep[][]{baranne1996}. \sophie\ is a cross-dispersed, stabilized \'echelle spectrograph on sky since 2006, dedicated to high-precision RV measurements \citep[][]{Perruchot-2008,Bouchy-2009}. Observations were taken in the fast-readout mode of the detector and in the high-resolution ($\lambda/\Delta\lambda \,=\,75 000  $) configuration of the spectrograph. To evaluate the sky or moonlight pollution, one of the optical fibers was placed on the sky while the other was used for the starlight. Depending on the star, data were recorded over time spans of 2 to 15 years. Most of the spectra have a typical signal-to-noise ratio (\snr) per pixel of 50 at a wavelength of 550 nm. Wavelength calibrations were performed at the beginning and end of each observing night, and approximately every two hours during the night. We observed the stars using \elodie\ and \sophie. However, for the analysis, we considered the data as coming from three different instruments: \elodie , \sophie, and \sophie+. \sophie+ is the result of an upgrade made in June 2011 consisting mainly in the addition of octagonal fibers to guide the light to the dispersive elements \citep[Sect. 1 of][]{Bouchy-2013}. Three of the stars were observed with \elodie , \sophie, and \sophie+, while the three remaining stars were only observed with \sophie+.

The \sophie\ pipeline \citep[][]{Bouchy-2009} was used for extracting the spectra and cross-correlating them with a numerical stellar mask. We first considered a G2 mask and incorporated all the spectral orders to produce cross-correlation functions (\ccf s). The \ccf s \  were fitted with Gaussians to derive the RVs \citep[][]{baranne1996, Pepe-2002}.
We then tested the effects of changing the spectral type of the numerical mask and/or removing some of the blue orders due to their low \snr. No significant changes were found, except for HD211403 and HD115954.
For HD211403, when we used a K5 instead of a G2 mask, we observed a significant decrease in the dispersion of the residuals of the fitted model (presented in \sect{planet}). This is surprising given the spectral type of the star (F7V). The analyses of the \rv s obtained with a K5, a G2, and a F0 mask provide consistent estimates of the system parameters, but with slightly smaller error bars when the K5 mask \rv s are used. However, as we do not have a convincing explanation for these slightly improved results, we use the \rv s obtained with the G2 mask for the rest of this paper.
For HD115954, the minimum of the dispersion of the residuals is obtained when we removed the four bluest and noisiest orders (thus keeping only orders 5 to 38).
The target \snr\ at 550\, nm for our observations is 50, but for some observations the observing conditions (in particular bad seeing or cloud coverage) did not allow us to reach this \snr . When the \snr\ obtained is less than 25, it implies that the observing conditions were particularly bad, and, in addition to a loss of precision, the extracted \rv\ often suffer a loss of accuracy. We thus reject the eight observations for which the \snr\ obtained is less than 25.
There are also 39 spectra with significant sky background pollution (moonlight contamination). We corrected for this using the spectrum of the sky background observed in the second optical fiber (fiber B) located 2\,arcmin away from the first optical fiber (fiber A) observing the star. The correction procedure is described in details in \citet{hebrard2008} and \citet{bonomo2010}. The amplitude of this correction for the 39 spectra affected varies in between 0.2 and 34.5 \ms .
Finally, to eliminate outliers, we removed any points lying beyond 9 -$\sigma$ of the RV residuals distribution.

The bisector span (\bis) was obtained following the approach of \citet[][]{queloz2001}. In the case of the fast rotator HD211403, we implemented the method of \citet[][]{Boisse-2011}. Both methods track the asymmetry in the line profile, but the approach by \citet[][]{Boisse-2011} is less sensitive to noise in the stellar line profile. The typical uncertainties in RVs are between 3 and 10 \ms\ for \sophie\ and $\sim$ 30 \ms\ for \elodie. We also quadratically added an uncertainty of 5 \ms\ to account for the poor scrambling properties to the exposures taken with \sophie\ before the June 2011 upgrade \citep{Hebrard-2016}. In addition, we also derived two activity indicators: the depth of the \halpha\ line \citep{Boisse-2011}, and of the Ca H\&K line through the $\textrm{logR'}_{\textrm{HK}}$ obtained directly from the SOPHIE DRS following the approach in \citet{boisse2010}. The measurements of the \rv , the full width at half maximum (\fwhm ) of the \ccf,  the \bis , the \logrhk, and the \halpha\ indicators are provided in \App{rvs}.

\section{Stellar characterization}\sectlabel{star}

The number of individual spectra for each star varies from 23 (HD95544) to 59 (HD80869). The S/N of the combined spectra for these stars is $\sim$ 300 at 550 nm.

\subsection{Spectroscopic logg, Teff, and [Fe/H]}\sectlabel{spectro}

The spectroscopic parameters were derived using our standard ``ARES+MOOG'' method \citep[for more details, see][]{sousa2014}. The spectral analysis is based on the excitation and ionization balance of iron abundance. The strengths of the absorption lines are consistently measured with the ARES v2 code \citep[][]{Sousa-2007, Sousa-2015b} and the abundances are derived in local thermodynamic equilibrium with the MOOG v2014 code \citep[][]{Sneden-1973}. For this step we used a grid of Kurucz ATLAS9 plane-parallel model atmospheres \citep[][]{Kurucz-1993}. The list of iron lines is the one taken from \citet[][]{Sousa-2008}, except for the stars with effective temperature below 5200 K where we used instead a more adequate line list for cooler stars \citep[][]{Tsantaki-2013}. This method has been applied in our previous spectroscopic studies of planet-hosts stars, which are all compiled in the Sweet-CAT catalog \citep[][]{Santos-2013,sousa2018}. The effective temperature (\teff ), surface gravity (\logg ) and iron metallicity (\feh ) values and errors bars obtained are provided in \tab{sysparam}.

The standard ARES+MOOG method described above is limited by stellar rotation ($> 10-15\ \kms$), because in this case the equivalent widths of the iron lines cannot be measured as precisely. Of our targets, only HD211403 shows such high stellar rotation. For this star, we applied the spectral synthesis technique based on the spectral package SME \citep[][]{valenti1996} and the methodology described in \citet[][]{tsantaki2014}. In particular, synthetic spectra are created for small regions around the iron lines and are compared with the observations under a $\chi^2$ minimization process, yielding \teff , \logg , \feh, and \vsini . This method is tested to preserve homogeneity with the equivalent width method, meaning that the parameters derived with both methods are on the same scale \citep[][]{tsantaki2014}.

\subsection{Stellar modeling: Mass and radius}\sectlabel{iso}

We modeled the planet-host stars using stellar evolutionary models generated from a 1D stellar evolution code -- Modules for Experiments
in Stellar Astrophysics (MESA; \citealt{Paxton1,Paxton2,Paxton3}). The physics used when constructing the stellar grid is similar to that in the GS98sta grid described in
section 3.1 of \citet{Nsamba}. We note that the stellar mass interval was extended to cover a mass range of M $\in$ [0.7 - 1.6] $\rm M_\odot$. In addition, for stars above 1.1 $\rm M_\odot$,
core overshoot is also included.
We used the optimization code AIMS (\citealp[Asteroseismic Inference on a Massive Scale;][]{Rendle}) to determine fundamental stellar properties, having adopted a number of observational constraints, namely, effective temperature, metallicity, and parallax-based luminosity.

We included systematic uncertainties of 59 K in effective temperature and 0.062 dex in metallicity that
arise from variation in the spectroscopic methods, as described by \citet{Torres2}.
The stellar luminosities (L) were calculated using the relation expressed as \citep{Pijpers}
\begin{equation}
 \rm log(L/L_\odot) = 4 + 0.4\,M_{\rm bol,\odot} - 2 \rm \log\pi[mas] - 0.4\,(V- A_v +BC_v)
 \label{1}
,\end{equation}
where $\pi$ is the parallax obtained from the Gaia DR2 data \citep{Gaia}, $\rm A_v$ is the extinction in the V band determined using Eqs. 19 and 20 of
\citet{Bailer}, V is the visual magnitude extracted from the Extended Hipparcos Compilation (XHIP; \citealt{Anderson}), and
$\rm BC_v$ is the bolometric correction determined using the \citet{Flower} polynomials expressed in terms of the effective temperature, later corrected by \citet{Torres1}.
A value of 4.73 mag for $\rm M_{\rm bol,\odot}$ is adopted in Eq.~\ref{1} since the \citet{Flower} polynomials are used to determine $\rm BC_v$.
This is because the bolometric magnitude of the Sun must be consistent with the zero point of $\rm BC_v$ so that the apparent brightness of the Sun is reproduced (see \citealt{Torres1}).
We note that $\rm A_v$ was obtained based on multiband photometry \citep{Bailer} since two of the stars in our sample (namely, HD 27969 and HD 211403) have low galactic latitudes and have a much lower true $\rm A_v$ than the full extinction of the Galaxy along their line of sight, as provided by commonly used dust maps such as MWDUST3\footnote{http://github.com/jobovy/mwdust} code \citep{Bovy}.

We note that AIMS combines a Markov chain Monte Carlo (MCMC) approach and Bayesian technique to explore the model parameter space and find models that have parameters comparable to the specified sets of observables.
The total $\chi^2$ combining the different observables in the optimization process takes the form
\begin{equation}
\chi_{\rm{total}} ^2 = \chi_{T_{\rm{eff}}} ^2 + \chi_{\rm{[Fe/H]}} ^2 + \chi_{\rm{L}} ^2 ~,
\label{chi-2}
\end{equation}
\noindent
where~
$\chi_{T_{\rm{eff}}} ^2 = \left(\frac{T_{\rm{eff}} ^{\rm(obs)} - T_{\rm{eff}} ^{\rm(mod)}}{\sigma (T_{\rm{eff}})}  \right)^2 $ ~,~
$\chi_{\rm{[Fe/H]}} ^2 = \left(\frac{{\rm{[Fe/H]}}^{\rm(obs)} - {\rm{[Fe/H]}}^{\rm(mod)}}{\sigma {(\rm{[Fe/H]})}}  \right)^2$ ~, ~ and
$\chi_{\rm{L}} ^2 = \left(\frac{\rm{L} ^{\rm(obs)} - \rm{L}^{\rm(mod)}}{\sigma (\rm{L})}  \right)^2 $~.

\noindent
The inferred stellar parameters and their corresponding uncertainties are obtained as the statistical mean and standard deviation of the posterior distributions.

The masses and radii of our six stars obtained through this method are presented in the stellar parameter section of \tab{sysparam}. Our six stars are consistent with main sequence stars with spectral types from F7 to G4. The stellar masses span from 0.98 to 1.20\,M$_{\sun}$ with relative precision from 4.5 to 6\,\%. The stellar radii span from 1.06 to 1.27\,R$_{\sun}$ with relative precision from 4 to 8\,\%.

\section{Planet characterization}\sectlabel{planet}

The generalized Lomb-Scargle periodograms \citep[\glsp s;][]{zechmeister2009} of the RV data sets of our 6 stars each exhibit at least one highly significant periodic signal (see \fig{timeseriesglsHD27969} to \figalt{timeseriesgls3}).
In \sect{plparam} through \sectref{HD211403}, we explore the planetary hypothesis for the origin of these signals. In \sect{validation}, we investigate other hypotheses and validate the planetary nature of the detected signals.

\subsection{The planetary model}\sectlabel{plparam}

Since our RV survey aims at discovering exoplanets, we first analyzed our six data sets under the planetary hypothesis.
As we see in \sect{validation}, this hypothesis is indeed the one favored by the data for all data sets.

We used part of the \texttt{radvel.kepler.rv\_drive} function of the Python package \texttt{radvel}\footlabel{opensourcesoftwares}{Several of the Python packages used for this work are publicly available on Github: \texttt{radvel} at \url{https://github.com/California-Planet-Search/radvel}, \texttt{emcee} at \url{https://github.com/dfm/emcee.}}
\citep[][]{fulton2018} to model the RV signal induced by the presence of a planet.
To ease the exploration of the parameter space, we used a parametrization of the model that minimizes the correlation between parameters \citep[modified from][]{eastman2013}: $P$ the orbital period, $t_{\mathrm{ic}}$ the planet's time of inferior conjunction, $e \cos \omega_*$ and $e \sin \omega_*$, where $e$ is the orbital eccentricity and $\omega_*$ is the stellar orbital argument of periastron, $K$ the RV semi-amplitude, and $v_0$ the systemic RV. To this set of parameters we added an additive jitter term ($\sigma_{\textit{inst}}$) for each instrument to account for underestimated uncertainties \citep[see][]{baluev2009}. Finally, we added a parameter for the shift of the RV zero point between the instruments ($\Delta\textrm{RV}$). The final list of model parameters is $P$, $t_{\mathrm{ic}}$, $e \cos \omega_*$, $e \sin \omega_*$, $K$, $v_0$, $\sigma_{\textit{inst}}$ and $\Delta\textrm{RV}$ (with as many $\sigma_{\textit{inst}}$ parameters as instruments and one fewer $\Delta\textrm{RV}$ parameter than the number of instruments).

To infer the best set of parameter values, we sampled the \post\ probability density function (\pdf) from the Bayesian inference framework \citep[e.g.,][]{gregory2005a}.
The \prior\ PDFs assumed for the parameters are given in \tab{priors} and described in \app{priors}. We chose physically motivated priors, not restricted by the indication provided by the \glsp , as is customary to do. This approach allows a blind search of the data to be performed. In all cases except for HD80869 (see \sect{HD80869}), it allows us to confirm the most significant peak of the \glsp\ as the planetary orbital period. For HD80869, the highest peak is actually an alias of the orbital period.

For the likelihood functions, we used multidimensional Gaussians, including the additive jitter terms ($\sigma_{\textit{inst}}$) as described by \citet{baluev2009}.
To locate the maximum of the \post\ \pdf\ (\textsc{map}) and infer reliable error bars for the parameters, we explored the parameter space using an affine-invariant ensemble sampler for \textsc{mcmc} thanks to the Python package \texttt{emcee}\footref{opensourcesoftwares} \citep[see][]{foreman-mackey2013}.
We adapted the number of walkers to the number of free parameters in our model. As a compromise between the speed and the efficiency of the exploration, we chose to use five times the number of free parameters for the number of walkers.

In each case, we performed an exploration with 50,000 iterations per walker with the initial values of the parameters drawn from the priors. Due to the wide prior on the orbital period, this exploration often (see sections \sectalt{HD27969} to \sectalt{HD211403}) identifies several local maxima of the \post\ \pdf . The traces of the walkers (plot of the values taken by each walker versus iterations for each parameter) show that not all walkers converged toward the same region of the parameter space. To identify the global maximum of the \post\ \pdf\ from this exploration, we plotted the histogram of the highest \post\ values identified by each walker. When several maxima are identified by the exploration, we can clearly see well separated peaks in this histogram. We then selected the walkers that belong to the peak with the highest \post\ values\footnote{Quantitatively, the walkers that belong to the highest peak are defined as all walkers whose highest \post\ values satisfy this criterion: $\ln(\mathrm{post})_{\mathrm{walker}} > \max\left(\ln(\mathrm{post})_{\mathrm{walker}}\right) - 5 * \textsc{mad}(\ln(\mathrm{post})_{\mathrm{walker}})$ where $\ln(\mathrm{post})_{\mathrm{walker}}$ is the natural logarithm of the highest \post\ value identified by a walker, $\max\left(\ln(\mathrm{post})_{\mathrm{walker}}\right)$ is the maximum of the logarithm of the \post\ identified by all walkers and \textsc{mad} is the median absolute deviation. We visually confirmed that this criterion corresponds to a peak in the histogram of $\ln(\mathrm{post})_{\mathrm{walker}}$
.}. For these walkers, we then identified and removed the burn-in phase with the Geweke algorithm \citep[see][]{geweke1992} and obtained the converged sample of iterations. From this sample, we used the median as estimator of the best value of each parameter. We used the 68\,\% confidence level interval estimated via the $16^{\textrm{th}}$ and $84^{\textrm{th}}$ percentiles of the converged chains as an estimator of the uncertainties.

In most cases, this exploration with 50,000 iterations per walker allowed more than 100,000 iterations to be obtained in the converged sample\footnote{In our analysis the typical correlation length of the chain is between 5 and 20, which means that we have 5 to 20 times fewer independent samples.} and we used these values and uncertainties as final results. When this is not the case, we performed a second exploration with 50,000 iterations per walker for which we drew the initial positions of the walkers from Gaussian distributions using the values and uncertainties derived from the first exploration as mean and standard deviation. We repeated the same treatment of the resulting chains to obtain the final parameter estimates.

The final estimates for our six systems are reported in \tab{sysparam}. The time series and \glsp\ of the \rv\ data, the best-fit models and their residuals are displayed in \app{timeseriesgls} (\fig{timeseriesglsHD27969} to \figalt{timeseriesgls3}). The phase-folded \rv\ curves are displayed in \fig{phasefold}.

Besides the model parameters, we also computed the values and uncertainties of secondary parameters. These parameters were computed from the model parameters and the stellar parameters (see \Sect{star}), but provide additional insights on the detected planets. The list of secondary parameters is: the time of periastron passage ($t_{\mathrm{p}}$), the orbital eccentricity ($e$), the stellar orbital argument of periastron ($\omega_*$), the minimum planetary mass ($\mpl \sin i_{\mathrm{p}}$),
the semimajor axis of the planetary orbit ($a$), the ratio of the semimajor axis to the stellar radius ($\frac{a}{R_*}$), the planetary equilibrium temperature ($T_{\textrm{eq}}$) assuming a geometric albedo of 0 and full energy redistribution over the planetary surface, and the incident flux on the planetary atmosphere ($F_{i}$). We computed the value of these parameters for each iteration of the converged iterations sample. This gave us chains of values for these secondary parameters and allowed us to compute the best values and uncertainties with the exact same procedure as for the model parameters. Some of these parameters rely on the stellar parameters ($M_{*}, R_{*}, T_{\mathrm{eff}}$). For these cases, we simulated chains for these parameters by randomly drawing values from normal distributions whose mean and standard deviations are set to the estimates provided by our stellar characterization analyzes (\sect{star}).

\begin{figure*}[!htb]
    \centering
    \subfloat[\textbf{HD27969}: rms of the residuals = 5.8 \ms\ (\sophiep)]{\includegraphics[width=0.446\linewidth]{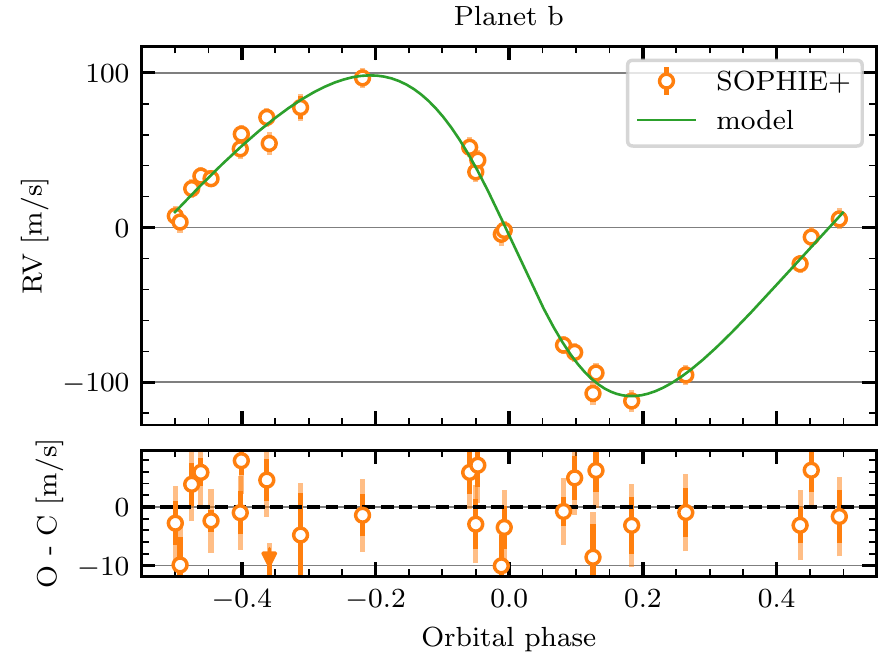}}
\hfil
    \subfloat[\textbf{HD80869}: rms of the residuals = 6.3, 6.3, 31.1 \ms\ (\sophiep, \sophie, \elodie)]{\includegraphics[width=0.446\linewidth]{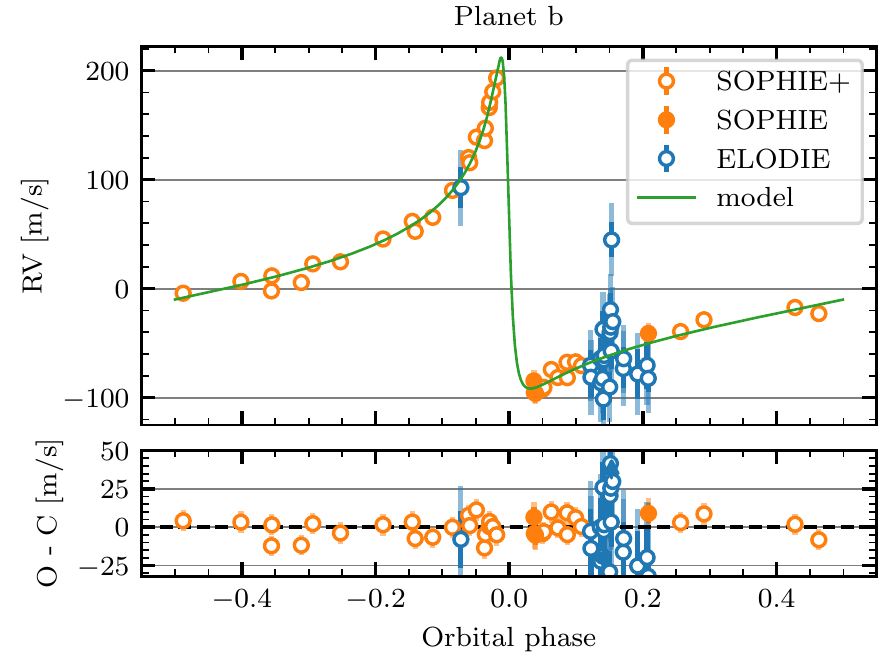}}

    \subfloat[\textbf{HD95544}: rms of the residuals = 4.4 \ms\ (\sophiep)]{\includegraphics[width=0.446\linewidth]{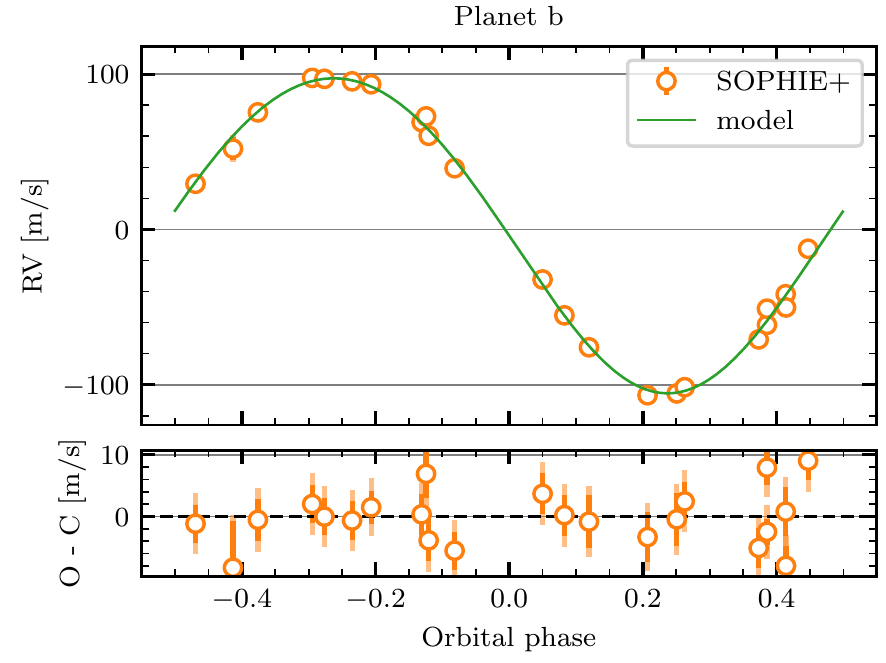}}
\hfil
    \subfloat[\textbf{HD109286}: rms of the residuals = 12.5 \ms\ (\sophiep)]{\includegraphics[width=0.446\linewidth]{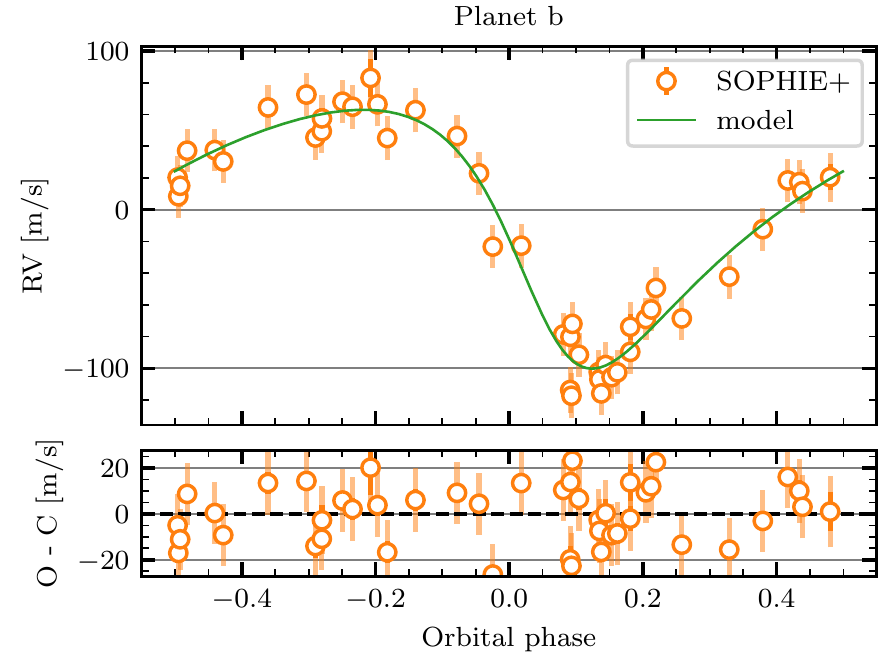}}
\hfil
    \subfloat[\textbf{HD115954}: rms of the residuals =  7.5, 5.3, 22.7 \ms\ (\sophiep, \sophie, \elodie)]{\includegraphics[width=0.446\linewidth]{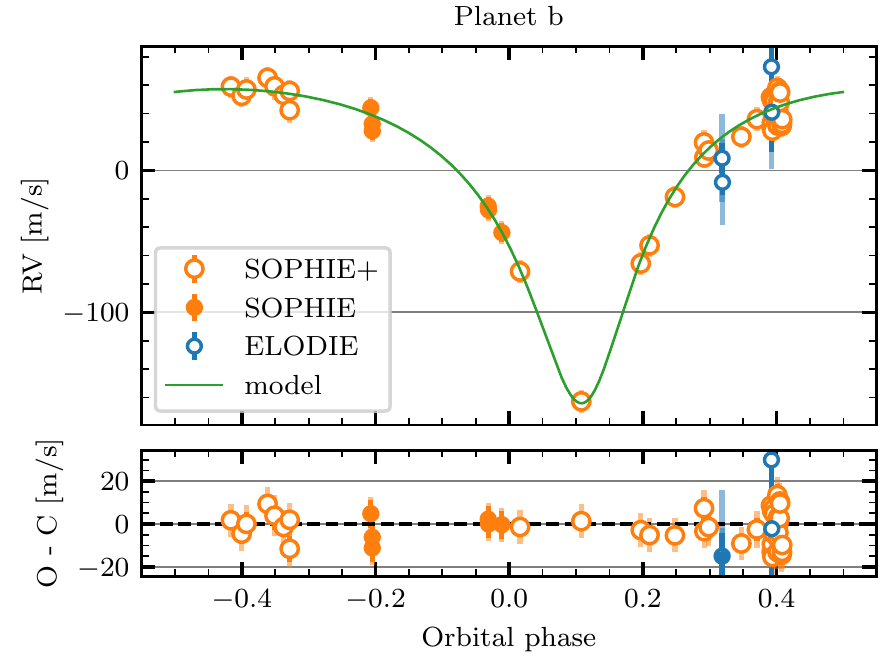}}
\hfil
    \subfloat[\textbf{HD211403}: rms of the residuals =  35.9, 37.9 and 76.6 \ms\ (\sophiep, \sophie, \elodie)]{\includegraphics[width=0.446\linewidth]{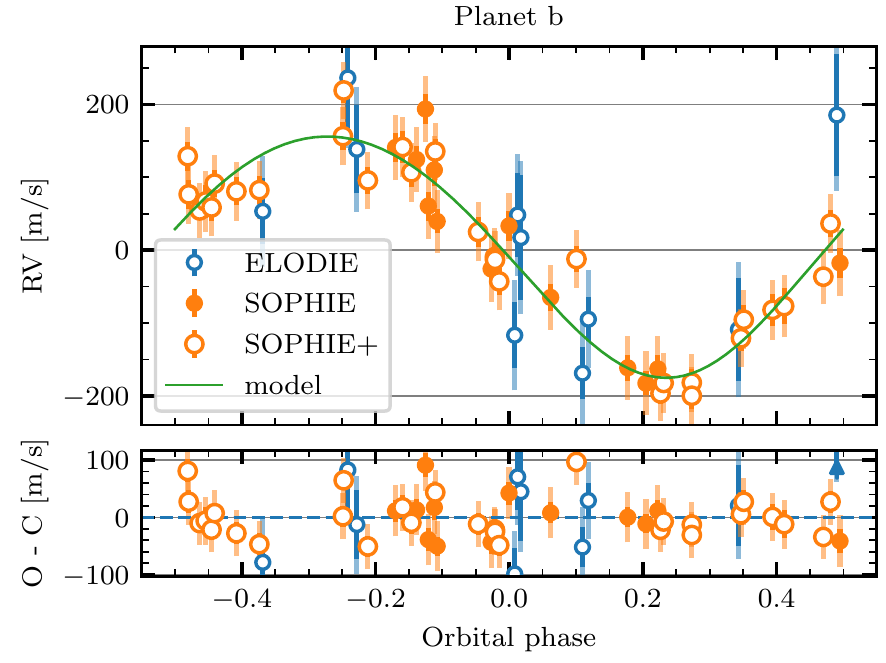}}

    \caption{\figlabel{phasefold}Phase-folded \rv\ curves for our six planets with the best-fit models. The color and filling of the points indicate the instrument used to acquire the data: empty blue for \elodie, filled orange for \sophie, and empty orange for \sophiep. The error bars provided with the \rv\ data are displayed with the same opacity and color as the points. The extended error bars computed with the fitted jitter parameters are displayed with a higher transparency. The best-fit model is represented with a green line. The \rv\ time series (not phase folded) are also displayed in \fig{timeseriesglsHD27969} to \figref{timeseriesgls3}.}
\end{figure*}

\subsection{HD27969}\sectlabel{HD27969}

For this system, we have 25 measurements from \sophiep. The \glsp\ of the data (\fig{timeseriesglsHD27969}) shows two peaks with a false alarm probability\footnote{The \fap\ levels are computed using the analytical relation described in \citet{zechmeister2009} when the Z-K normalization is used.}
(\fap) lower than 0.1\,\% around 1 and 600 days. We performed one exploration with 50,000 iterations per walkers. It identified six local maxima. The difference of the logarithm of the \post\ \pdf\ values ($\Delta \ln(\mathrm{post})$) of the five lowest local maxima compared to the highest one are $\Delta \ln(\mathrm{post}) > 21$. In order to interpret these values of $\Delta \ln(\mathrm{post})$, it is convenient to see each local maxima as a separate model. We wanted to understand if one of these models is significantly better to explain the data. To do that, we used $\Delta \ln(\mathrm{post})$ values as a proxy for the Bayes factor (the ratio of the likelihood of the two models being compared). This allowed us to interpret the values of $\Delta \ln(\mathrm{post})$ following \citet[][]{KassRaftery1995} or \citet[][]{jeffreys1998}. We thus considered that $\Delta \ln(\mathrm{post}) > 5$ implies that there is strong evidence that the model with the highest \post\ \pdf\ is indeed the favored model. We acknowledge that $\Delta \ln(\mathrm{post})$ is not an accurate estimator of the Bayes factor. We would need to integrate over the support of each maximum to provide a more accurate estimator. However, with values above 21 for a threshold of 5 on a logarithmic scale, $\Delta \ln(\mathrm{post})$ appears to be sufficient in this case. After selecting the global maximum and removing the burn-in phase, we obtained 280,000 converged iterations.

The best-fit Keplerian points toward a giant planet with a minimum mass of $4.80_{-0.23}^{+0.24}\,\mjup$, an orbital period of $654.5_{-5.8}^{+5.7}$ days and a significant eccentricity of  $0.182_{-0.019}^{+0.019}$.
We note that the time span of the data just covers one orbital period. An \rv\ trend of instrumental or astrophysical origin would thus be difficult to differentiate from the Keplerian model if it is smaller than the semi-amplitude. As a consequence, even if no trend is observed in the residuals of the model (see \fig{timeseriesglsHD27969}), a small trend could be absorbed by the Keplerian model and lead to a slight over-estimation of the semi-amplitude and eccentricity. The dispersion of the residuals of the best-fit model is 5.8\,\ms, which represents 1.5 times the average error bar of the \sophiep\ \rv s, indicating that there are probably no other significant signals in the data. Furthermore, there is no significant peak in the \glsp\ of the residuals (see \fig{timeseriesglsHD27969}). The weighted average of the \logrhk\ time series is -5.3, confirming that the star is quiet.

\subsection{HD80869}\sectlabel{HD80869}

For this system, we have 59 measurements: 22 from \elodie, 4 from \sophie, and 33 from \sophiep.  The \glsp\ of the data (\fig{timeseriesglsHD80869}) shows nine peaks with an \fap\ lower than 0.1\,\% around 0.3, 0.5, 1, 140, 330, 430, 600, 900, and 2000 days. We performed a first exploration with 50,000 iterations per walkers, which identified five local maxima. The difference of the logarithm of the \post\ \pdf\ values of the four lowest local maxima compared to the highest one are $\Delta \ln(\mathrm{post}) > 185$. After selecting the global maximum and removing the burn-in phase, we obtained 70,000 converged iterations. We thus performed a second exploration with 50,000 iterations per walker to better explore the global maximum and obtain 2,520,000 converged iterations.

The best-fit Keplerian points to a giant planet with a minimum mass of $4.86_{-0.29}^{+0.65}\,\mjup$, an orbital period of $1711.7_{-9.6}^{+9.3}$ days, and a significant eccentricity of $0.862_{-0.018}^{+0.028}$.
The time span of the observations covers slightly more than three orbital periods and no trend is observed in the residuals of the model (see \fig{timeseriesglsHD80869}). The dispersion of the residuals of the best-fit model is 6.3, 6.3, and 31.1\,\ms, which represents 2.5, 1, and 1.8 times the average error bar for \sophiep, \sophie, and \elodie, respectively. The dispersion of the residuals compared to the average error bar of \sophiep\ \rv s might indicate that there are other signals in the data. The \glsp\ of the residuals shows a peak at around 11 days (see \fig{timeseriesglsHD80869}), but its low significance ($\sim 10\,\%$ \fap ) does not allow us to further investigate its origin. The weighted average of the \logrhk\ time series is -5.1, indicating a relatively quiet star.

HD80869 is the only case (within our six systems), for which the highest peak in the \glsp\ of the \rv s is does not correspond to the estimated orbital period of the planet (see \fig{timeseriesglsHD80869}). The highest peak is actually the fourth harmonic of orbital period. This is due to a combination of harmonics and aliases. As shown in the middle-left panel of \fig{timeseriesglsHD80869}, the high eccentricity of the orbit implies that a relatively high fraction of the power spectral density is distributed in the harmonics. This, coupled with the fact that the fourth harmonic also coincides with an alias of the orbital period and its three first harmonics (see the lower-left panel of \fig{timeseriesglsHD80869}) explains why the highest peak corresponds to a fifth of the orbital period instead of to the orbital period itself.

\subsection{HD95544}\sectlabel{HD95544}

For this system, we have 23 measurements from \sophiep. The \glsp\ of the data (see \fig{timeseriesgls2}-a) shows two peaks with an \fap\ lower than 0.1\,\% around 1 and 2000 days. We performed a first exploration with 50,000 iterations per walker, which identified eight local maxima. The difference of the logarithm of the \post\ \pdf\ values of the seven lowest local maxima compared to the highest one are $\Delta \ln(\mathrm{post}) > 85$. After selecting the global maximum and removing the burn-in phase, we obtained 80,000 converged iterations. We thus performed a second exploration with 50,000 iterations per walker to better explore the global maximum and obtained 2,520,000 converged iterations.

The best-fit Keplerian points to a giant planet with a minimum mass of $6.84_{-0.31}^{+0.31}\,\mjup$, an orbital period of $2172_{-21}^{+23}$ days, and a low-significance eccentricity of $0.043_{-0.016}^{+0.017}$.
As for HD27969, the time span of the data just covers one orbital period. No trend is observed in the residuals of the model (see \fig{timeseriesgls2}-a), but a relatively small trend could thus be absorbed by the Keplerian model and lead to a slight over-estimation of the derived semi-amplitude and eccentricity. The dispersion of the residuals of the best-fit model is 4.4\,\ms, which represents 1.3 times the average error bar of the \sophiep\ \rv s, indicating that there is probably no other significant signal in the data. The \glsp\ of the residuals (see \fig{timeseriesgls2}-a) does not show any significant peak. The weighted average of the \logrhk\ time series is -5.2, confirming that the star is quiet.

\subsection{HD109286}\sectlabel{HD109286}

For this system, we have 45 measurements from \sophiep. The \glsp\ of the data (see \fig{timeseriesgls2}-b) shows five peaks with an \fap\ lower than 0.1\,\% around 0.23, 0.33, 0.5, 1 and 500 days. We performed one exploration with 50,000 iterations per walker. It identified five local maxima. The difference of the logarithm of the \post\ \pdf\ values of the five lowest local maxima compared to the highest one are $\Delta \ln(\mathrm{post}) > 100$. After selecting the global maximum and removing the burn-in phase, we obtained 1,240,000 converged iterations.

The best-fit Keplerian points toward a giant planet with a minimum mass of $2.99_{-0.15}^{+0.15}\,\mjup$, an orbital period of $520.1_{-2.3}^{+2.3}$ days and a significant eccentricity of $0.338_{-0.035}^{+0.034}$.
The time span of the observations covers around four orbital periods. A trend was initially observed in the residuals of the best-fit model, which led us to fit a linear trend to the data (see \tab{sysparam} and \fig{timeseriesgls2}-b). The significant slope detected (see \tab{sysparam}) does not display any correlation with the other parameters of the model. It could be a sign of an additional longer period body in the system (see \sect{astrometry}), or a long-period activity signal like a magnetic activity cycle (see \sect{stelact}). The dispersion of the residuals of the best-fit model (which includes the \rv\ trend) is 12.5 \ms, which represents 2.7 times the average error bar of \sophiep\ data. This could indicate the presence of other signals in the data, like stellar activity. However, the \glsp\ of the residuals (see \fig{timeseriesgls2}-b) does not show any significant peak. The weighted average of the \logrhk\ time series is -4.45, confirming that the star is active and that the extra dispersion observed in the residuals could be due to stellar activity.

\subsection{HD115954}\sectlabel{HD115954}

For this system, we have 49 measurements: 4 from \elodie, 6 from \sophie, 39 from \sophiep. The \glsp\ of the data (see \fig{timeseriesgls3}-a) shows three peaks with an \fap\ lower than 0.1\,\% around 0.5, 1 and 3000 days. We performed one exploration with 50,000 iterations per walker. It identified three local maxima. The difference of the logarithm of the \post\ \pdf\ values of the two lowest local maxima compared to the highest one are $\Delta \ln(\mathrm{post}) > 50$. After selecting the global maximum and removing the burn-in phase, we obtained 1,175,000 converged iterations.

The best-fit Keplerian points toward a giant planet with a minimum mass of $8.29_{-0.58}^{+0.74}\,\mjup$, an orbital period of $3700_{-390}^{+1500}$ days and a significant eccentricity of $0.487_{-0.041}^{+0.095}$.
As for HD27969 and HD95544, the time span of the data covers just one orbital period, but this time a trend was initially observed in the residuals of the model, which led us to fit a linear trend to the data (see \tab{sysparam} and \fig{timeseriesgls3}-a). The fitted slope coefficient is not significant (see \tab{sysparam}), but displays a clear correlation with the orbital eccentricity and the \rv\ offset between \sophie\ and \sophiep , and a slight correlation with the \rv\ semi-amplitude.
The dispersion of the residuals of the best-fit model is 7.5, 5.3, and 22.7\,\ms, which represents 1.6, 0.8, and 1.5 times the average error bar for \sophiep, \sophie\ and \elodie, respectively.
This indicates that there is probably no other significant signal in the data. The \glsp\ of the residuals (see \fig{timeseriesgls3}-a) does not show any significant peak. The weighted average of the \logrhk\ time series is -5.1 confirming that the star is quiet.

\subsection{HD211403}\sectlabel{HD211403}

For this system, we have 52 measurements: 10 from \elodie, 13 from \sophie, 29 from \sophiep. The \glsp\ of the data (see \fig{timeseriesgls3}-b) shows three peaks with an \fap\ lower than 0.1\,\% around 0.5, 1 and 220 days. We performed one exploration with 50,000 iterations per walker. It identified five local maxima. The difference of the logarithm of the \post\ \pdf\ values of the four lowest local maxima compared to the highest one are $\Delta \ln(\mathrm{post}) > 25$. After selecting the global maximum and removing the burn-in phase, we obtained 375,000 converged iterations.

The best-fit Keplerian points toward a giant planet with a minimum mass of $5.54_{-0.38}^{+0.39}\,\mjup$, an orbital period of $223.8_{-0.41}^{+0.41}$ days and a low-significance eccentricity of $0.084_{-0.044}^{+0.057}$.
The time span of the observations covers around 23 orbital periods. No trend is observed in the residuals of the best-fit model (see \fig{timeseriesgls3}-b). The dispersion of the residuals of the best-fit model is 35.9, 37.9 and 76.6\,\ms, which represents 1.9, 2.0, and 1.3 times the average error bar for \sophiep, \sophie,\ and \elodie,\ respectively.
This indicates that there is probably no other clearly significant signal in the data. The \glsp\ of the residuals (see \fig{timeseriesgls3}-b) does not show any significant peak. The weighted average of the \logrhk\ time series is -4.6, which indicates a relatively active star.

The estimated \rv\ offset of $-219\,\ms$ between the \elodie\ and \sophiep\ instruments is relatively large compared to the value obtained for HD80869 and HD115954 (see \tab{sysparam}). An abnormally high \rv\ offset could be the sign of an \rv\ drift and of an additional companion. We thus looked at the \rv\ offset measurements published by \citet{Boisse-2012} and \citet{Kiefer2019a}\footnote{We want to caution the reader interested in the \rv\ offset between \elodie\ and \sophie . \citet{Kiefer2019a} mentioned that the expected range is between 50 and 100\,\ms according to \citet{Bouchy-2013}. This reference is probably erroneous and should be \citet{Boisse-2012}. Furthermore, the 50-100\,\ms\ range is valid only for \rv s computed with a G2 mask.} to provide a broader context. In particular, Fig A.1 of \citet{Boisse-2012} shows the \rv\ offset between \elodie\ and \sophie\ for a sample of stable stars. The authors showed that when using a G2 mask, as we do for HD211403, the measured \rv\ offset is typically in the range 0 to -120\,\ms . This \rv\ offset could thus be the sign of an \rv\ drift and of the presence of a companion at a larger orbital period.

\subsection{Validation of the planetary nature}\sectlabel{validation}

Periodic signals similar to the one produced by a planet orbiting the target star can be produced by spectroscopic binaries (SBs), contaminating spectroscopic binaries (CSBs), hierarchical triple systems (HTSs), and stellar activity \citep[see for example][]{Santerne2015, queloz2001}.

The first hypothesis to explore is the SB, where the system observed is composed of two stellar and gravitationally bound objects. As far as the identification of such systems is concerned, two cases need to be considered. The first case is the SB2. The magnitudes of both stars are sufficiently close to be able to observe two sets of lines in the observed spectra. A visual inspection of the spectrum also enables the detection of some cases of HTS and CSB, where the apparent magnitude of the target star is similar to that of one of the other stars involved in these scenarios. The inspection of our spectra enables us to reject these scenarios.

The second case is the SB1. In this scenario, the second star is too faint compared to the target star to enable the identification of two sets of lines. However, the gravitational pull of the stellar companion induces a large RV amplitude. The analysis of such systems under the planetary hypothesis, as performed in \sect{plparam}, leads to mass estimates that indicate a stellar companion. Assuming that the orbital inclination is not too low (the system is not face-on), the mass estimates provided in \tab{sysparam} enable us to reject this hypothesis for most orbital inclinations.

\subsubsection{False positive indicators}\sectlabel{fp}

For the two remaining false positive cases, which imply a gravitationally bound system (i.e., HTS and CSB), the target star is not the one receiving the gravitational pull that we would detect.
As shown by \citet[][]{Santerne2015}, the \ccf\ of the star receiving the pull is blended with the one of the target star and deforms it. These deformations induce an \rv\ signal when the \ccf\ is fitted with a Gaussian profile. Such a scenario thus implies a deformation of the \ccf\ profile, which can be captured by the \bis\ and FWHM of the \ccf. If the detected \rv\ variations originate from these deformations the \rv\ will correlate with the \bis\ and/or the FWHM.

\begin{table}[!htb]
\small
\caption{\tablabel{cor} 99.7\% confidence interval for the correlation coefficients between RV and BS, FWHM, and $\textrm{logR'}_{\textrm{HK}}$.}
\centering
\begin{tabular}{lcccc}
\hline
Star         & \rv\ vs \bis   & \rv\ vs \fwhm  & \rv\ vs $\log(\textrm{R'}_{\textrm{HK}})$ \\
\hline\\[-5pt]

HD27969  & $[-0.67, 0.33]$  & $[-0.08, 0.82]^{*}$       & $[-0.50, 0.51]$        \\
HD80869  & $[-0.46, 0.48]$  & $[-0.51, 0.41]$       & $[-0.17, 0.71]$    \\
HD95544  & $[-0.70, 0.26]$  & $[-0.81, 0.06]^{*}$       & $[-0.63, 0.43]$        \\
HD109286 & $[-0.55, 0.22]$  & $[-0.39, 0.40]$       & $[-0.52, 0.40]$    \\
HD115954 & $[-0.24, 0.54]$  & $[-0.54, 0.21]$       & $[-0.48, 0.35]$    \\
HD211403 & $[-0.47, 0.35]$  & $[-0.36, 0.47]$       & $[-0.59, 0.19]$    \\

\hline \hline
\end{tabular}
\tablefoot{These confidence intervals have been computed using the method described in \citet[][]{figueira2016a}.\\
$^{*}$ indicates that, the correlation coefficient is compatible with 0 (no correlation) according to the 99.7\,\% confidence interval, but not according to the 95\,\% confidence interval.}
\end{table}

In \tab{cor}, we present estimates of the correlation coefficient between the RV, \bis\ and FWHM for our six systems obtained with the method described in \citet[][]{figueira2016a}. All cases are compatible with no correlation (0) according to the 99.7\% confidence interval, but two of them are not according to the 95\% confidence interval (see the asterisks in \tab{cor}).
This analysis put some constraints on the CSB and HTS scenarios, but does not completely exclude them. It is indeed possible that the precision of the RV, \bis\ and FWHM measurements does not allow a significant detection of an existing correlation. Marginal detections (like the two that we mentioned) can be triggered due to the relatively small sample size ($\sim 25$ to $60$ measurements in these cases).

\newcommand{\maxfraction}[1]{\mbox{$\mathrm{Max}(\frac{#1}{RV})$}}

\begin{table}[!htb]
\small
\caption{\tablabel{disp} Analysis of the dispersion of the \bis\ and \fwhm.}
\centering
\begin{tabular}{lcccc}
\hline\\[-5pt]
Star         & \dispratio{BS}  & \maxfraction{BS}  & \dispratio{FWHM}  & \maxfraction{FWHM}\\
             &                 & $\mathrm{[\%]}$   &                   & $\mathrm{[\%]}$\\
\hline\\[-5pt]

HD27969      & $1.34 \pm 0.21$ &               8.1 & $1.50 \pm 0.24$   &                22 \\
HD80869      & $0.88 \pm 0.15$ &               8.3 & $1.69 \pm 0.35$   &                39 \\
HD95544      & $1.11 \pm 0.18$ &               5.2 & $1.29 \pm 0.21$   &                15 \\
HD109286     & $1.31 \pm 0.16$ &               6.5 & $2.13 \pm 0.26$   &                26 \\
HD115954     & $1.38 \pm 0.16$ &              10.4 & $1.32 \pm 0.15$   &                25 \\
HD211403     & $1.90 \pm 0.21$ &                30 & $1.26 \pm 0.14$   &                49 \\

\hline \hline
\end{tabular}
\tablefoot{ $<\sigma_{\mathrm{X}}>$
indicates the average error bar on the individual measurements of X.\\
\maxfraction{X} is the maximum fraction of the observed rv amplitude that X can have without producing
a \dispratio{X} ratio larger than one at 3 sigma; see \sect{fp} for more details on the interpretation.
}
\end{table}

To address this point, we performed a simple dispersion analysis as described in \citet[][]{demangeon2018}. It consists of computing the ratio of the dispersion over the average measurement error for the \bis\ and FWHM measurements (see results in \tab{disp}). In short, in the CSB and HTS scenarios, because the RV signal is produced by the deformation of the CCF, both \bis\ and FWHM have to exhibit a dispersion whose amplitude is larger than their average error bar ($\dispratio{X} > 1$). The values of \dispratio{X} for our data sets are provided in \tab{disp}. If produced by a CSB or an HTS, this extra dispersion is equal to a fraction of the RV dispersion and should correlate with it. As described by \citet[][]{Santerne2015}, the value of this fraction depends on the characteristics of the CBS and HTS systems: magnitude ratio, mean RV separation, FWHM of the stellar components and spectral types.
In \tab{disp}, we also provide the maximum fraction of the RV dispersion that the FWHM or \bis\ can have without triggering a 3 sigma detection of an extra dispersion. This number allows us to understand how constraining this analysis is for each case.
The dispersion analysis also complements the correlation analysis, since it accounts for measurement uncertainties. It indicates whether the dispersion of the measurements requires more than pure measurement uncertainty to be explained. If it is not the case, the correlation analysis cannot provide a reliable correlation detection. Marginal detections, like the two presented in \tab{cor}, can then safely be ignored. If extra dispersion is detected, then the correlation analysis should be able to tell if it is correlated to the RV dispersion and thus if we can reject the planetary hypothesis in favor of the CSB or HTS hypothesis. \tab{disp} displays only one case, the \fwhm\ of HD109286, where the dispersion of the measurements cannot be explained solely by the measurement uncertainties: $\dispratio{FWHM} = 2.13 \pm 0.26,$ which is thus different from 1 at 4.3 sigma. As this extra dispersion does not correlate with the RV, we can still exclude the hypothesis that the source of this dispersion is also the source of the Keplerian-like signal observed in the \rv\ and thus reject the CSB or HTS hypothesis. As we discuss in more detail in \sect{stelact}, the origin of this dispersion is probably stellar activity.

\subsubsection{Astrometry}\sectlabel{astrometry}

To investigate the CSB or HTS hypothesis even further, we inspected the Gaia Archive database\footnote{http://gea.esac.esa.int/archive/} \citep[][]{gaiacollab2016, gaiacollab2018} to find possible astrometric motions for these systems. The presence of an astrometric motion could show either an inclination different from edge-on or the presence of an unseen long-period companion
\citep[][]{Kiefer2019a, Kiefer2019b}.

The six sources presented in this paper were all observed by the Gaia space telescope with data published in Data Release (DR) 1 and 2. Somewhat significant excess noise \citep[$>$0.5 mas; ][]{Kiefer2019a} was found in the DR1 for only two systems, HD27969 ($\epsilon_\text{DR1}$=0.62\,mas) and HD109286 ($\epsilon_\text{DR1}$=0.88\,mas). Nevertheless, these excess noises hardly stand out compared to the distribution of astrometric excess noise published in the DR1 for all primary sources monitored with Gaia, of which a majority are single stars, with a median at 0.45 mas and a 90th percentile at 0.85 mas.
The excess noise measured for HD109286 is slightly larger than this limit, and could therefore be real. Moreover, the instrumental and photon noise are minimized about the Gaia magnitude of this star G $\sim$ 8.6 \citep[][]{Lindegren2018}.

On the other hand, inspecting the DR2 archive, we did not find large deviations from a good astrometric fit with 5 parameters, with $\chi^2$=282 for 158 degrees of freedom. This is a reduced $\chi^2$ of 1.78, or a unit weight error UWE=1.33. According to \citep[][]{Lindegren2018}, this is close to the median UWE at this magnitude of about 1.4. Therefore, no significant excess astrometric noise seems to be detected in DR2.
If a (likely small) part of the DR1 excess noise measured for HD109286 is real, then the fact it is not observed in DR2 implies that it could be due to an unseen long-period companion. The astrometric reflex motion of the star due to this companion may be partly fitted in DR2, which would not be the case in DR1. Indeed, the Tycho-2 and Hipparcos-2 positioning from 24 years ago was taken into account in DR1 but not in DR2 \citep[][]{Lindegren2016, Lindegren2018}. This hypothesis of an unseen long-period companion is reinforced by our detection of a linear \rv\ trend (see \sect{HD109286}).

We also examined the Hipparcos Intermediate Astrometry Data using the methods described in \citep[][]{Sahlmann2016} to investigate whether they can put constraints on the systems' parameters, and in particular on the masses of the companions. \Tab{astro} lists the target names and the basic parameters of the Hipparcos observations relevant for the astrometric analysis. The solution type (Sol. Type) indicates the astrometric model adopted by the new reduction. For the standard five-parameter solution it is “5.” The parameter $N_{\textrm{orb}}$ represents the number of orbital periods covered by the Hipparcos observation time span. $N_{\textrm{Hip}}$ is the number of astrometric measurements with a median precision of $\sigma_\lambda$. The last column in \tab{astro} shows the maximum companion mass ($M_{2,\textrm{max}}$) that would be compatible with a non-detection in the Hipparcos astrometry.
We do not detect significant orbital motion in the astrometry for any of these sources, but we determine upper mass limits of 0.14, 12.02, and 0.6\,$M_{\sun}$ for the companions of HD27969, HD109286, and HD211403, respectively.

 \begin{table}[!htb]
 \small
 \caption{\tablabel{astro} Parameters of the Hipparcos astrometric observations.}
 \centering
 \begin{tabular}{lcccccc}
 \hline\\[-5pt]
 Target & HIP & Sol. & $N_{orb}$ & $\sigma_{\lambda}$  & $N_{Hip}$ & $M_{2,\mathrm{max}}$\\
        &     & type &           & (mas)                 &           &   ($M_\sun$)\\
 \hline\\[-5pt]

  HD27969 & 20753  & 5 & 1.7 & 4.2 & 172 & 0.14 \\
  HD80869 & 46022 & 5 & 0.5 & 3.8 & 112 & N/A \\
  HD95544 & 54203 & 5 & 0.5 & 5.9 & 160 & N/A \\
  HD109286 & 61298 & 5 & 2.3 & 7.9 & 103 & 12.02 \\
  HD115954 & 65042 & 5 & 0.3 & 6.6 & 59 & N/A \\
  HD211403 & 109876 & 5 & 5.0 & 4.9 & 140 & 0.60 \\

 \hline \hline
 \end{tabular}
 \end{table}

For our six targets, the Gaia and Hipparcos data thus agree with the planetary origin of the signal observed. They do not provide any indication that it could instead be produced by an SB, a CSB, or an HTS. Even if we cannot exclude all configurations of SB, CSB, or HTS, the simplest explanation is the planetary one. One false positive scenario, however, remains to be explored: stellar activity.

\subsubsection{Investigating the stellar activity hypothesis}\sectlabel{stelact}

Stellar activity, in particular spots and plages, produces deformations of the CCF profile and intensity variations in stellar lines that are particularly sensitive to activity, like the flux variation at the core of the Ca II H\&K lines measured by the \logrhk\ indicator \citep[e.g.,][]{duncan1991}. It can thus produce quasi-periodic \bis , \fwhm , \logrhk\ and \rv\ variations due to the intrinsic periodicity of the activity cycles or stellar rotation \citep[e.g.,][]{queloz2001}. In such a case, we expect an extra dispersion in the \fwhm\ and/or \bis\ and/or \logrhk\ measurements, as observed for HD109286 (see \tab{disp}). If the signal observed in the RV data of HD109286 was due to stellar activity, we would also expect a correlation between the \fwhm\ and the \rv\ measurements \citep[e.g.,][]{dumusque2014a}. This was not observed (see \tab{cor}) for HD109286 or any of other targets. This indicates that stellar activity is unlikely to be the source of the six RV signals that we detect. However, it could explain the extra dispersion of the \fwhm\ detected in the data of HD109286.
This explanation is supported by the measurements of the \logrhk, a well-known activity indicator \citep[e.g.,][]{Wright2004}. HD109286 exhibits the highest \logrhk\ with an average value of -4.45.

As the periods of the \rv\ signals detected are quite large (from 224 to 3700 days), we must mention stellar magnetic cycles as another potential source of false positives. However, observations of \rv signals associated with magnetic cycles have so far constrained their amplitude to a few tens of \ms \citep[e.g., ][]{gomesdasilva2012,lovis2011a}, one order of magnitude smaller than any of the signals that we detect.

To confirm the rejection of the stellar activity hypothesis (whether coming from a magnetic cycle or not), in \app{actinglsp}, we show the \glsp\ of the \bis , \fwhm\ and \logrhk\ time series (\fig{actinHD27969} to \figref{actinHD211403}). If the observed signals were due to stellar activity, we would expect to observe a peak at the same period in these \glsp s. This is not the case for any of our six planets. These \glsp\ could also show peaks at the stellar rotation period, but no significant peak (with \fap\ level below 10\,\%) is observed in the any of \bis , \fwhm\ or \logrhk\ \glsp s for any of our six systems.

To conclude, the six periodic signals discovered in the RV observations of HD27969, HD80869, HD95544, HD109286, HD115954 and HD211403 are all of planetary origin to the best of our knowledge.

\subsubsection{Search for additional planetary signals}\sectlabel{morepl}

Kima\footnote{\url{https://github.com/j-faria/kima}} is a Python/C++ open-source package developed for the detection of exoplanets with RV data~\citep{Faria2018_kima}. It makes use of the diffusive nested sampling algorithm~\citep{Brewer2011_DNS} to infer the number of planets present in a set of RV measurements.
 To do this, kima calculates the evidence for a model with a given number of planets ($N_p$). In our work $N_p$ is set as a free parameter between 0 and 2, and we evaluate the posterior distribution for $N_p$ to determine the number of planets detected. This means that to claim the detection of $N_p$ planets, the probability of $N_p$ planets needs to be at least 150 times greater than the probability of $N_{p-1}$ planets~\citep{KassRaftery1995}.

 The analyses of the \rv\ series of our six targets conclude that in all cases the best model is a model with only one planet. Such a confirmation is particularly important for planets with high orbital eccentricity. A highly eccentric orbit can indeed mimic the signal from two planets on close to circular orbits \citep[][]{wittenmyer2019a,wittenmyer2013}.

Furthermore, for HD109286 and HD115954, our model also included an \rv trend that could be due to an additional body with an even longer orbital period \citep[e.g.,][]{wittenmyer2019} in the system, or to stellar activity (like the signature of a magnetic cycle).
Finally, Kima also provides the parameters of the detected planets. All parameters are compatible within 1 sigma with the ones obtained by the analysis described in \sect{plparam}.

\section{Discussion}\sectlabel{discussion}

In this paper we have presented the detection of six giant planets with minimum masses ranging from 2.99 to 8.29 \mjup\ and long orbital periods ranging from 223.7 to 3700 days ($\sim 10$\,years). We now discuss the importance of these planets in the context of the known exoplanet population.

\subsection{The importance of the cold and eccentric giant planet population}

\Fig{planetsdistrib} presents our six planets in the mass-period-eccentricity diagram along with the known exoplanet population. Our six planets belong to a relatively populated part of the distribution: the cold giant planets. According to \href{https://exoplanetarchive.ipac.caltech.edu/index.html}{NASA Exoplanet Archive} \citep{akeson2013}, we currently know of 354 planets with a mass larger than 0.2\,\mjup\ and an orbital period above 100 days\footnote{On top of the mass and period lower limits provided, we only considered exoplanets whose mass and orbital period are measured with a relative precision better than 50\,\%.}.
Within this population, HD 80869 b stands out as the planet with the seventh most eccentric orbit\footnote{According to the NASA exoplanet archive, HD 80869 b is the planet with the sixth most eccentric orbit, but the archive does not yet take into account the recently updated orbital parameters of HD 76920 b \citep{bergmann2021}.}, and HD 115954 b and HD 109286 b also have eccentricities above the median eccentricity (0.198) of the known exoplanet population\footnote{We again only considered planets whose mass and orbital period are measured with a relative precision better than 50\,\%.}.

\begin{figure}[!htb]
    \resizebox{\hsize}{!}{\includegraphics[]{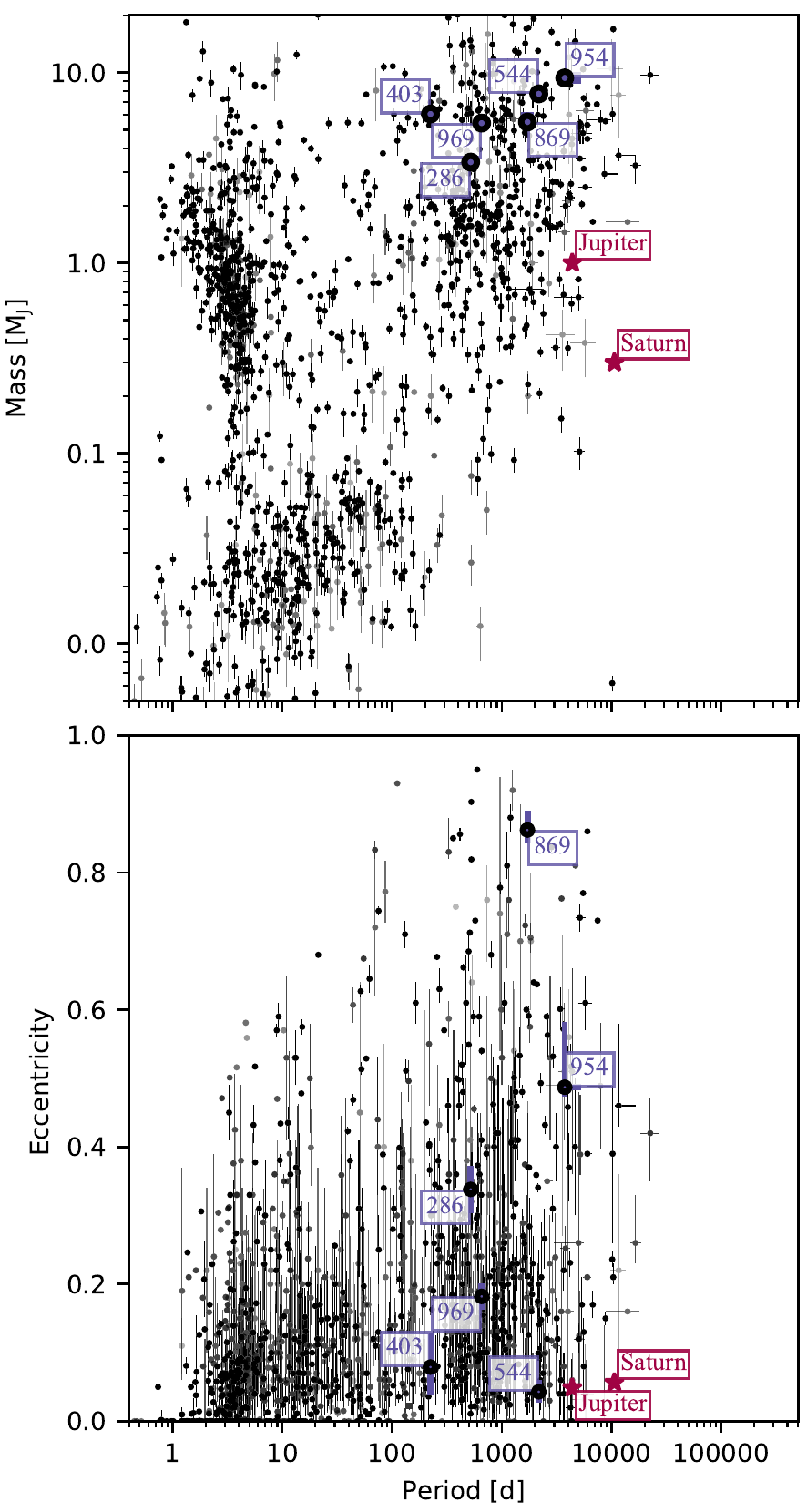}}
    \caption{\figlabel{planetsdistrib}Mass versus orbital period distribution (top) and orbital eccentricity versus orbital period distribution (bottom) of the known exoplanet population according to \href{https://exoplanetarchive.ipac.caltech.edu/index.html}{NASA Exoplanet Archive}. We only display exoplanets with a relative mass and orbital period precision better than 50\,\%. The six exoplanets announced in this paper are marked in blue with a thick circle, and the last three digits of their host star names are displayed (for example ``869'' for HD80869 b). The red stars indicate Solar System planets for reference.
    }
\end{figure}

Reaching a statistical understanding of the eccentricity distribution of the exoplanet population, and in particular its high tail, is essential for constraining planetary formation and migration models \citep[e.g.,][]{chatterjee2008,juric2008,bowler2020}. Eccentric planets are not a direct output of planet formation from core-accretion in a protoplanetary disk \citep[see for example][]{lin1997}. The interaction between the planet and the disk usually damps the eccentricity. Eccentric orbits thus require an additional process.
Three such processes are discussed in the literature: disk cavity migration, planet-planet scattering, and Kozai-Lidov perturbations.
Disk cavity migration \citep[e.g.,][]{goldreich2003,debras2021} is a special case of giant planet-disk tidal interaction that promotes eccentricity growth instead of damping it. Eccentricity damping relies on the balance between waves produced by two kinds of resonances: the Lindblad resonances, which excite the eccentricity, and the corotation resonances, which damp it \citep[e.g.,][]{papaloizou2001}. When a planet is more massive than $\sim 10\,\mjup$, it opens a cavity in the disk. If this cavity is wide enough to encompass the principal Lindblad resonances and the first-order corotation resonances, the planet's eccentricity is controlled by the next most important resonances: the first-order Lindblad resonances. This process can explain eccentricities of up to $\sim 0.4$.
The planet-planet scattering scenario \citep{rasio1996,weidenschilling1996,lin1997} relies on close encounters between two planets. The gravitational interaction between the two bodies can send them into eccentric orbits. If such an event arises after the dissipation of the protoplanetary disk, and the eccentric orbit does not trigger other close encounters with other planets or efficient tidal dissipation with the parent star, the planet can remain on its eccentric orbit.
The Kozai-Lidov perturbation \citep{kozai1962,lidov1962} also requires the presence of another massive body, a star or a massive substellar companion, but this time not located in the plane of the protoplanetary disk.
In such a case, the difference of orbital inclination between the planet and this third body can excite both the orbital inclination and the eccentricity of the planet.
These last two scenarios are often used to explain the existence of hot Jupiters \citep[see for example][]{winter2020,teyssandier2019}. Hot Jupiters would represent the fraction of eccentric giant planets whose eccentricity led to strong tidal interactions with the parent star, a dissipation of the angular momentum, and a circularization of the planet's orbit \citep[e.g.,][]{ogilvie2014}. The process or processes that explain the presence and characteristics of hot Jupiters must also explain the properties of the eccentric cold Jupiter population to which several of our newly discovered planets belong.

\subsection{The small planet-cold Jupiter relation}

A large fraction of the known planets belong to multiple systems\footnote{According to NASA exoplanet archive, currently 43\,\% of the known planets belong to a multiple system.}, and it is not rare for an apparently single-planet system to actually harbor other undetected planets \citep[][]{sandford2019}. We have seen that our data sets do not support the presence of additional planets in any of our six systems (\sect{morepl}). However, the observational strategy of our RV program is not favorable to multi-planetary system discoveries. The low cadence and relatively low RV precision of our data sets are well adapted for the detection of cold Jupiters, but not for the detection of super-Earths or sub-Neptunes on shorter orbits.

From the analysis of planets detected with both RV and transit photometry surveys, \citet{zhu2018} showed that there is a correlation between the presence of small short-period planets (planets with mass and radius between those of Earth and Neptune) and the presence of giant long-period\footnote{\citet{zhu2018} define a "long period" as an orbital period above one year.}  planets within the same system. More specifically, they conclude that cold Jupiters are almost certainly ($\sim 90\,\%$) accompanied by small planets. These results are in agreement with other independent studies \citep{bryan2016,bryan2019} and indicate that our six stars are likely to host small planets that are undetected in our data sets. Three out of the six cold Jupiters discussed in this paper have an orbital eccentricity higher than 0.3. We can expect a cold Jupiter with a high orbital eccentricity to compromise the stability of the inner planetary system and impact the probability of finding a small inner planet \citep{pu2018,mustill2017}. It is relevant at this point to mention that there is currently only one system known to host at least one small inner planet and a cold Jupiter with an orbital eccentricity above 0.8 \citep{santos2016}. It is thus not common, but also not impossible, for a system hosting a cold Jupiter with such a highly eccentric orbit to host a small inner planet. \citet{zhu2018} did not directly study this question, but they approached it indirectly by comparing the multiplicity of inner-planetary systems with and without a known cold Jupiter. As cold Jupiters are known to have eccentric orbits \citep[see for example Fig 2 of ][]{zhu2018}, if the presence of a cold Jupiter reduces the multiplicity of small inner planetary systems, the eccentricity of the orbit of the cold Jupiter can be proposed as an explanation, even if it is not the only possible explanation. In their sample, \citet{zhu2018} found that the average number of small inner planets is 2.5 for systems without a cold Jupiter and drops to 1.4 for systems that also host a cold Jupiter.

We will thus follow up on these six stars with a higher cadence and a higher \rv\ precision to detect additional small planets in these systems.

\begin{acknowledgements}

% Mine
O.D.S.D.~is supported in the form of work contract (DL 57/2016/CP1364/CT0004) funded by national funds through Funda\c{c}\~{a}o para a Ci\^{e}ncia e a Tecnologia (FCT).

% FCT
This work was supported by Fundação para a Ciência e a Tecnologia (FCT) and Fundo Europeu de Desenvolvimento Regional (FEDER) via COMPETE2020 through the research grants UIDB/04434/2020, UIDP/04434/2020, PTDC/FIS-AST/32113/2017 \& POCI-01-0145-FEDER-032113, PTDC/FIS-AST/28953/2017 \& POCI-01-0145-FEDER-028953.

% Vincent
This work has been carried out in part within the framework of the NCCR PlanetS supported by the Swiss National Science Foundation. This project has received funding from the European Research Council (ERC) under the European Union's Horizon 2020 research and innovation programme (project {\sc Spice Dune}; grant agreement No 947634).

% PNP remerciements :
I.B., G.H, X.D. received funding from the French Programme National de Physique Stellaire (PNPS) and the Programme National de Planétologie (PNP) of CNRS (INSU).

%Benard
B. N acknowledges postdoctoral funding from the Alexander von Humboldt Foundation.

% Tiago
T.L.C. is also supported by FCT in the form of a work contract (CEECIND/00476/2018).

%Melissa
M.J.H. acknowledges support from ANID – Millennium Science Initiative – ICN12 009.

% Benard
B.N. acknowledges postdoctoral funding from the Alexander von Humboldt Foundation and "Branco Weiss fellowship Science in Society" through the SEISMIC stellar interior physics group.

% Only catalogs
This work made use of Simbad, Vizier and exoplanet.eu.

% Open source softwares
Most of the analyses presented in this paper were performed using the Python language (version 3.5) available at \url{http://www.python.org} and several scientific packages: Numpy \citep{vanderwalt2011}, Scipy \citep{virtanen2020},  Pandas \citep{mckinney2010}, Ipython \citep{perez2007}, Astropy \citep{astropycollaboration2013, astropycollaboration2018} and Matplotlib \citep{hunter2007}.

\end{acknowledgements}

\bibliographystyle{aa}
\bibliography{bibliography}

%\pagebreak
\begin{appendix}

\section{Prior distributions}\applabel{priors}

\begin{table}[!htb]
\tiny
\caption{\tablabel{priors}Priors of Bayesian MCMC analysis.}
\raggedright
\begin{tabular}{lc}
\hline
Parameter & Prior \\
\hline\\[-5pt]
$P$ & $\mathrm{JP}\left[P: \mathcal{J}(1, 1.5 \Delta\mathrm{t}_{\rv})\,\textrm{days}, \phi_{\textrm{ic}}: \mathcal{U}(0, 1)\right]$\\
${t_{\textrm{ic}}}$ & $\mathrm{JP}\left[P: \mathcal{J}(1, 1.5 \Delta\mathrm{t}_{\rv})\,\textrm{days}, \phi_{\textrm{ic}}: \mathcal{U}(0, 1)\right]$\\
$K$ & $\mathcal{U}\left[0, 2\cdot \textrm{max}\left\{\mathrm{p2p}(\rv_{\mathrm{inst}}\right\}\right]$\\
$e\cos \omega_*$ & $\mathrm{JP}\left[e: \beta(0.867, 3.03), \omega_*: \mathcal{U}(-\pi, \pi)\right]$\\
$e\sin \omega_*$ & $\mathrm{JP}\left[e: \beta(0.867, 3.03), \omega_*: \mathcal{U}(-\pi, \pi)\right]$\\
$v0_{\sophiep}$ & $\mathcal{N}\left[\textrm{med}(\rv_{\sophiep}), \textrm{std}(\rv_{\sophiep})\right]$\\
$\Delta\textrm{\rv}_{\sophie/\sophiep}$ & $\mathcal{N}[\textrm{med}(\rv_{\sophie}) - \textrm{med}(\rv_{\sophiep}),$\\
& $\sqrt{\textrm{var}(\rv_{\sophie}) + \textrm{var}(\rv_{\sophiep})}]$\\
$\Delta\textrm{\rv}_{\elodie/\sophiep}$ & $\mathcal{N}[\textrm{med}(\rv_{\elodie}) - \textrm{med}(\rv_{\sophiep}),$\\
& $\sqrt{\textrm{var}(\rv_{\elodie}) + \textrm{var}(\rv_{\sophiep})}]$\\
$\sigma_{\sophiep}$ & $\mathcal{U}\left[0, 5 \cdot \textrm{med}(\sigma_{\rv_{\sophiep}})\right]$\\
$\sigma_{\sophie}$ &  $\mathcal{U}\left[0, 5 \cdot \textrm{med}(\sigma_{\rv_{\sophie}})\right]$\\
$\sigma_{\elodie}$ &  $\mathcal{U}\left[0, 5 \cdot \textrm{med}(\sigma_{\rv_{\elodie}})\right]$\\
\hline \hline
\end{tabular}
\tablefoot{$\mathcal{U}[vmin, vmax]$, $\mathcal{J}[vmin, vmax]$, $\mathcal{N}[\textrm{mean}, \textrm{std}],$ and $\beta(a, b)$ stand for uniform, Jeffreys, normal, and beta probability distributions, respectively. For the uniform and Jeffreys distributions, $vmin$ and $vmax$ are the minimum and maximum values. For the normal distribution, mean is self-explanatory and std is the standard deviation. For the beta distribution, a and b are the 2 shape parameters of the beta distribution. $\Delta\mathrm{t}_{\rv}$ designate the time span of the \rv\ observations (considering all instruments). med used in several priors is the abbreviation of median. Similarly, var is the abbreviation of variance. $\textrm{max}\left\{\mathrm{p2p}(\rv_{\mathrm{inst}})\right\}$ is the maximum of the pic-to-pic values computed for each instrument separately. $\mathrm{JP}$ stands for joint prior. These joint priors are described in more details in the text of \app{priors}}
\end{table}

\Tab{priors} lists the priors used for all the parameters of the model in the Bayesian analysis. $P$ and ${t_{\textrm{ic}}}$ on one side and $e\cos \omega_*$ and $e\sin \omega_*$ on the other are affect joint prior probabilities.
$P$ and ${t_{\textrm{ic}}}$ are affected by the joint prior $\mathrm{JP}\left(P: \mathcal{J}(1, 1.5 \Delta\mathrm{t}_{\rv})\,\textrm{days}, \phi_{\textrm{ic}}: \mathcal{U}(0, 1)\right)$.
In practice, this means that $P$ and ${t_{\textrm{ic}}}$ are used to compute the phase of inferior conjunction $\phi_{\textrm{ic}}$ defined as
$$\phi_{\textrm{ic}} = \frac{{t_{\textrm{ic}}} - t_{\textrm{ref}}}{P},$$
where $t_{\textrm{ref}}$ is defined as the time of the first measurement obtained with \sophiep\ floored to the closest integer. Then individual priors are affected to $P$ and $\phi_{\textrm{ic}}$: A Jeffreys prior between 0.1 day and 1.5 times the time span of the observations for $P$ and a uniform prior between 0 and 1 for $\phi_{\textrm{ic}}$.
The joint prior on $P$ and ${t_{\textrm{ic}}}$ can thus be written as

\begin{eqnarray*}
\mathrm{JP}\left(P: \mathcal{J}(1, 1.5 \Delta\mathrm{t}_{\rv})\,\textrm{days}, \phi_{\textrm{ic}}: \mathcal{U}(0, 1)\right) = &\\
f(P, t_{\textrm{ic}})\,dP\,dt_{\textrm{ic}} =
g(P, \phi_{\textrm{ic}})\,dP\,d\phi_{\textrm{ic}} = &\\
= \mathcal{J}(P | 1, 1.5 \Delta\mathrm{t}_{\rv}) . \mathcal{U}(\phi_{\textrm{ic}} | 0, 1) \,dP\,d\phi_{\textrm{ic}.}
\end{eqnarray*}
Here, $e\cos \omega_*$ and $e\sin \omega_*$ are affected by the joint prior $\mathrm{JP}\left(e: \beta(0.867, 3.03), \omega_*: \mathcal{U}(-\pi, \pi)\right)$. This means that we compute the eccentricity ($e$) and the argument of periastron of the stellar orbit ($\omega_*$) and affect a beta prior with shape parameters a and b equal to 0.867 and 3.03, respectively, to $e$ \citep[as suggested by][]{kipping2013b}, and a uniform prior between $-\pi$ and $\pi$ to $\omega_*$. The joint prior on $e\cos \omega_*$ and $e\sin \omega_*$ can thus be written as

\begin{eqnarray*}
\mathrm{JP}\left(e: \beta(0.867, 3.03), \omega_*: \mathcal{U}(-\pi, \pi)\right) = &\\
f'(e\cos \omega_*, e\sin \omega_*)\,de\cos \omega_*\,de\sin \omega_* =
g'(e, \omega_*)\,de\,d\omega_* = &\\
= \beta(e | 0.867, 3.03) . \mathcal{U}(\omega_* | -\pi, \pi) \,de\,d\omega_*.
\end{eqnarray*}

\section{\rv\ time series and generalized Lomb-Scargle periodograms}\applabel{timeseriesgls}

 We show in \fig{timeseriesglsHD27969} to \figref{timeseriesgls3} the \rv\ time series and the \glsp\ of the \rv\ data, the planetary model, the residuals and the window function. These figures allow a visual assessment of the quality of the fitted model and the presence of trends and additional signals in the residuals.

\begin{figure*}[!htb]
    \centering
    \includegraphics[]{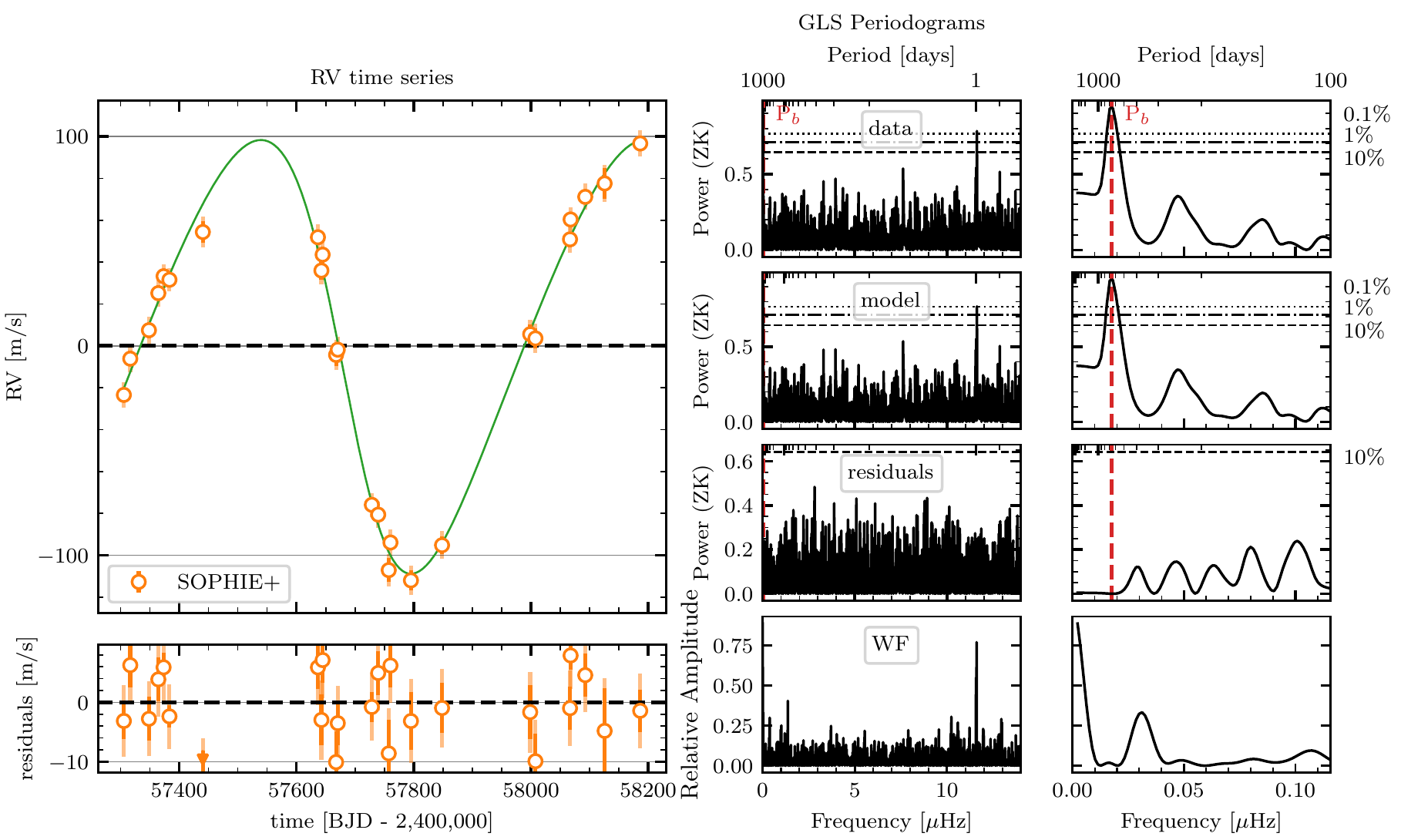}
    \caption{\figlabel{timeseriesglsHD27969}Radial velocities, best-fit model, and residuals for the HD27969 system. (Left-top) \rv\ time series and best-fit model, from which the systemic velocity and the offsets between instruments have been removed. (Left-bottom) Time series of the residuals of the best-fit model. The color and filling of the points indicate the instrument used to acquire the data: empty blue for \elodie, filled orange for \sophie, and empty orange for \sophiep. The error bars provided with the \rv\ data are displayed with the same opacity and color as the points. The extended error bars computed with the fitted additive jitter parameters are displayed with a higher transparency. (Middle, top) \glsp\ of the \rv\ time series, the best-fit model sampled at the same times as the \rv s (middle, second from the top), the residuals (middle, third from top), and the window function (middle, bottom). (Right) \glsp s zoomed-in around the period of the detected planet. The best-fit orbital period of the planet is marked by a vertical dashed red line. The horizontal dotted, dash-dotted, and dashed black lines correspond to 0.1, 1, and 10\,\% \fap\ \citep{zechmeister2009} levels, respectively.}
\end{figure*}

\begin{figure*}[!htb]
    \centering
    \includegraphics[]{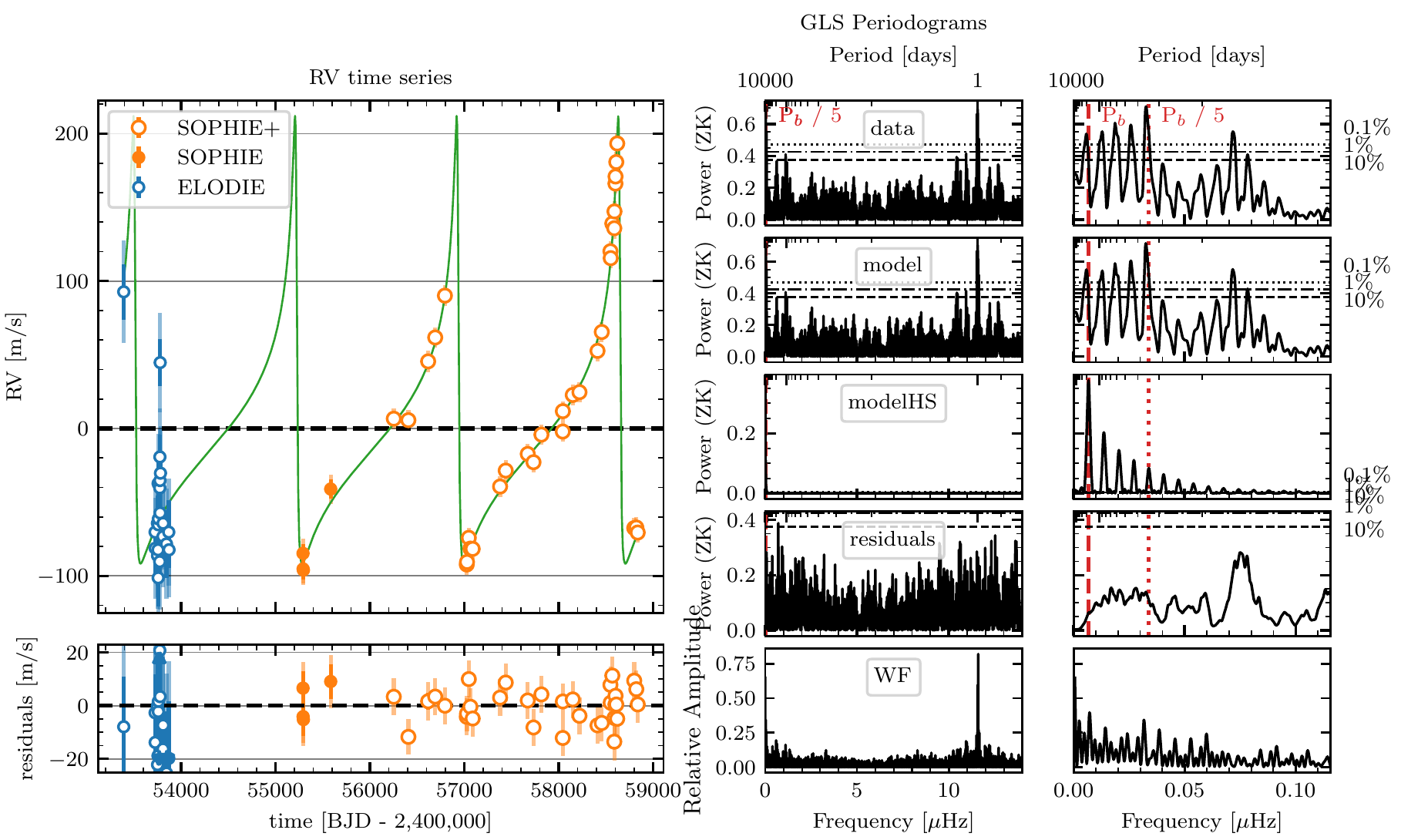}
    \caption{\figlabel{timeseriesglsHD80869}Radial velocities, best-fit model, and residuals for the HD80869 system. This figure is structured and generated in the same way as \fig{timeseriesglsHD27969} (see the caption of that figure for more details). The main difference is that in this case we add, in the third row, the \glsp\ of the planetary model sampled at 10,000 times, evenly spread over the time span of our observations (modelHS). This allows us to visualize the harmonic content of this highly eccentric orbit and better understand the \glsp\ of our \rv\ data.}
\end{figure*}

\begin{figure*}[!htb]
    \centering
    \subfloat[HD95544]{\includegraphics[]{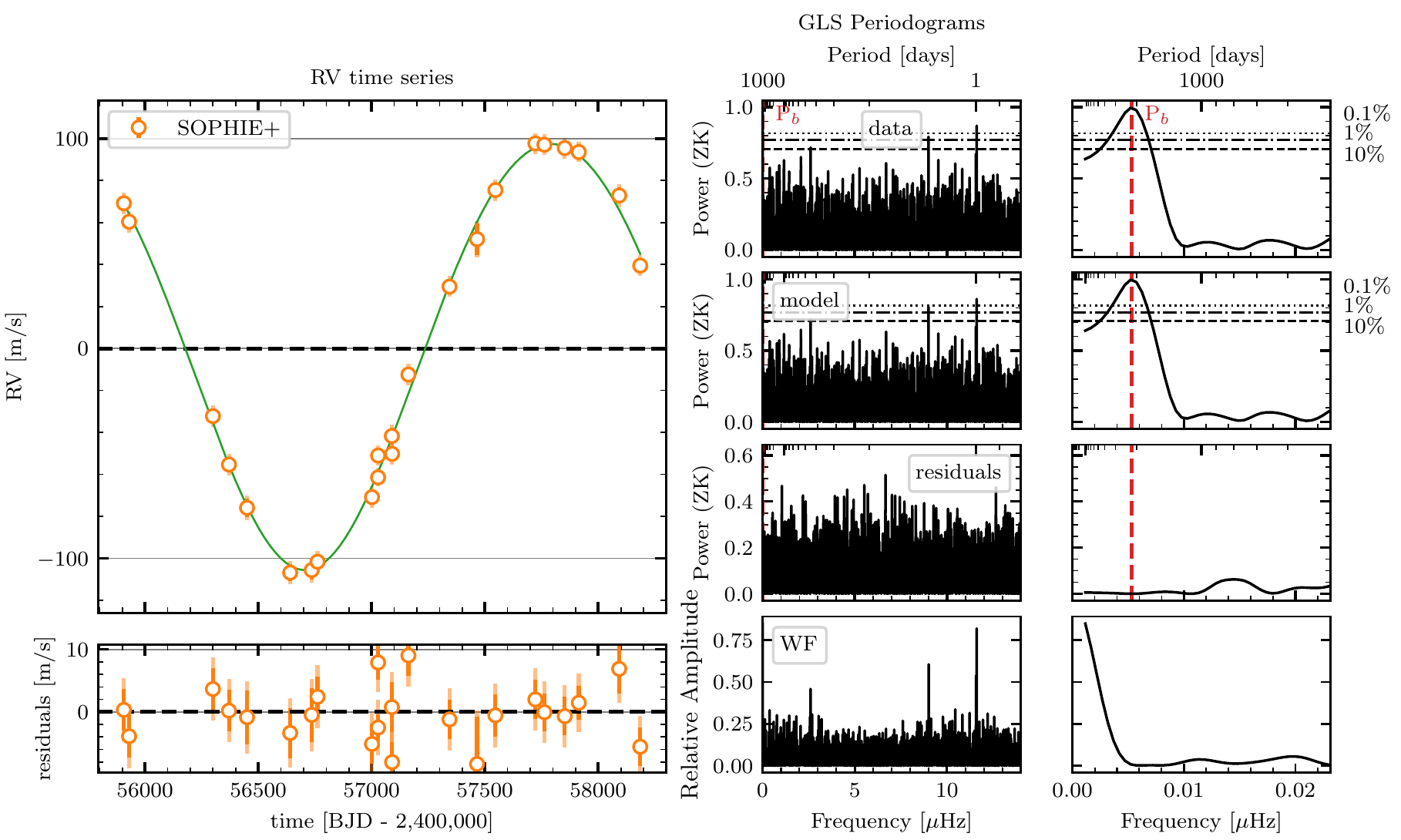}}
\hfil
    \subfloat[HD109286]{\includegraphics[]{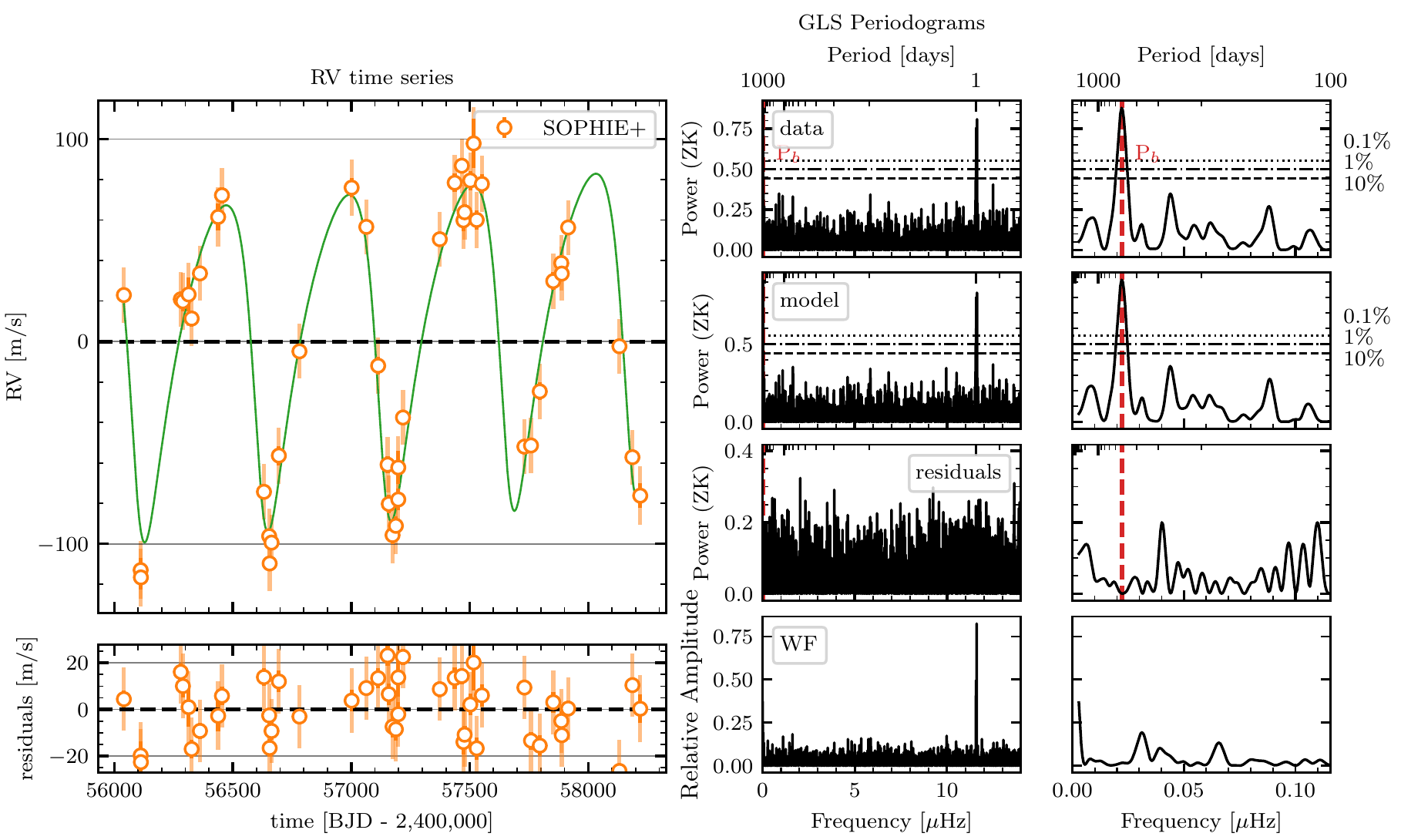}}
    \caption{\figlabel{timeseriesgls2}Radial velocities, best-fit model, and residuals for the HD95544 (a) and HD109286 (b) systems. This figure is structured and generated in the exact same way as \fig{timeseriesglsHD27969} (see the caption of that figure for more details).}
\end{figure*}

\begin{figure*}[!htb]
    \centering
    \subfloat[HD115954]{\includegraphics[]{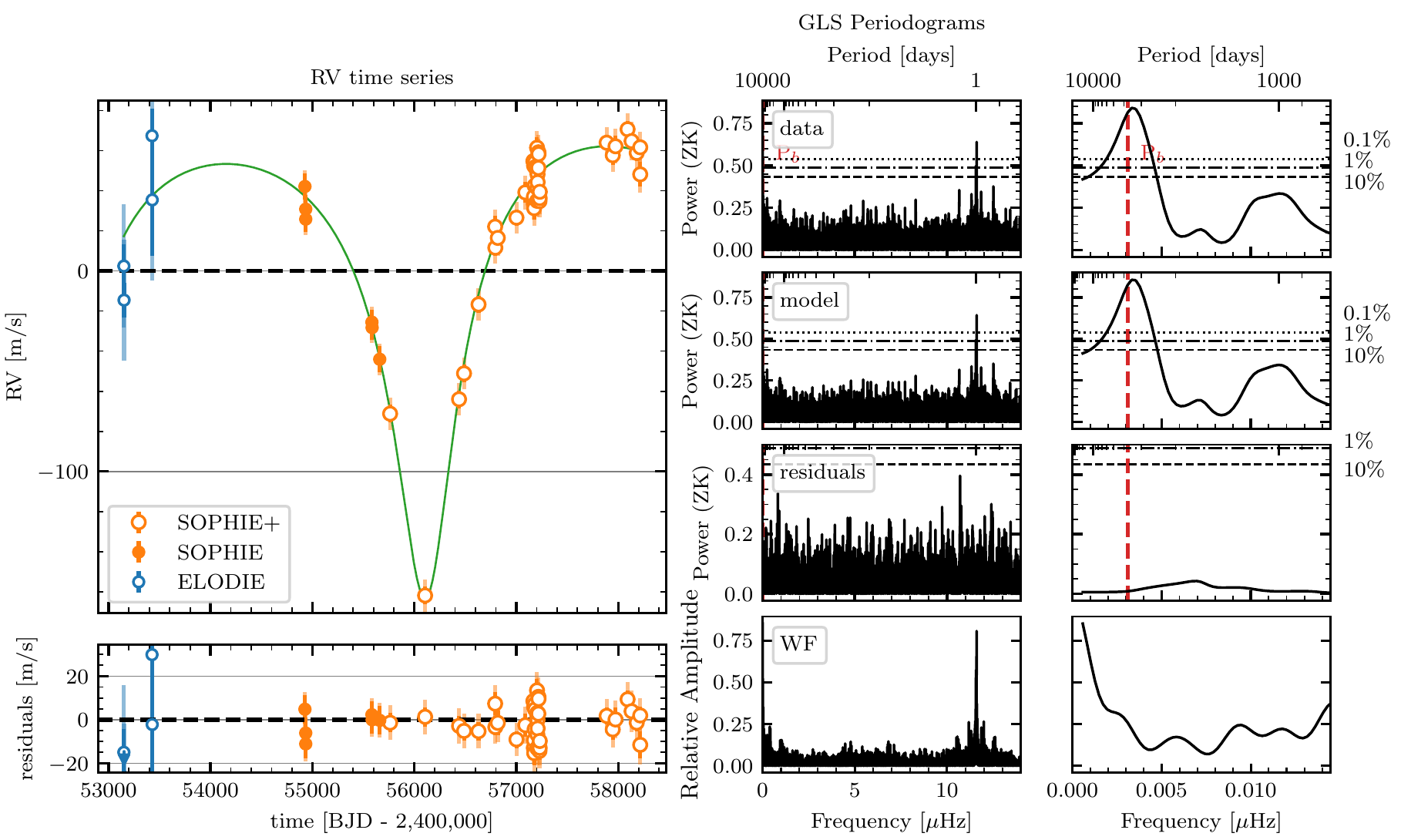}}
\hfil
    \subfloat[HD211403]{\includegraphics[]{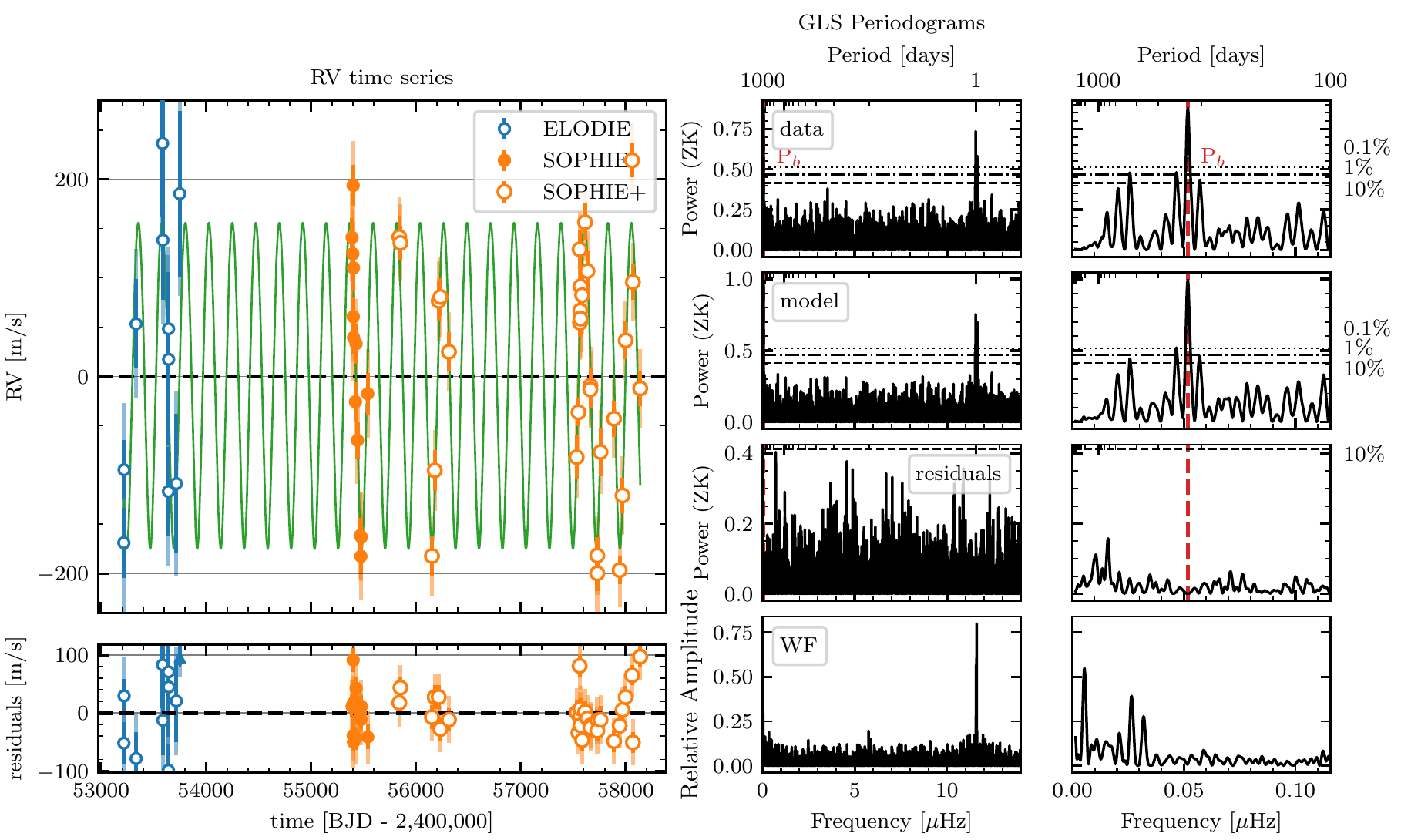}}
    \caption{\figlabel{timeseriesgls3}Radial velocities, best-fit model, and residuals for the HD115954 (a) and HD211403 (b) systems. This figure is structured and generated in the exact same way as \fig{timeseriesglsHD27969} (see the caption of that figure for more details).}
\end{figure*}

\section{\glsp\ of the activity indicators}\applabel{actinglsp}

\fig{actinHD27969} to \figref{actinHD211403} show the \glsp\ of the \rv , \fwhm , \bis\ and \logrhk\ data collected for our six systems and their associated window function. The frequency and period ranges are chosen to visualize the region surrounding the observed planetary periods and check that the \glsp\ of the \fwhm , \bis\ and \logrhk\ data do not show a significant peak at the same period. These frequency and period ranges are not always suited to visualizing shorter periods and assessing the presence of peaks associated with the stellar rotation period. However, we inspected these \glsp\ at shorter periods and do not report any peak with \fap\ levels below 10\,$\%$.

\begin{figure}[!htb]
    \centering
    \subfloat[HD27969 - \sophiep]{\includegraphics[]{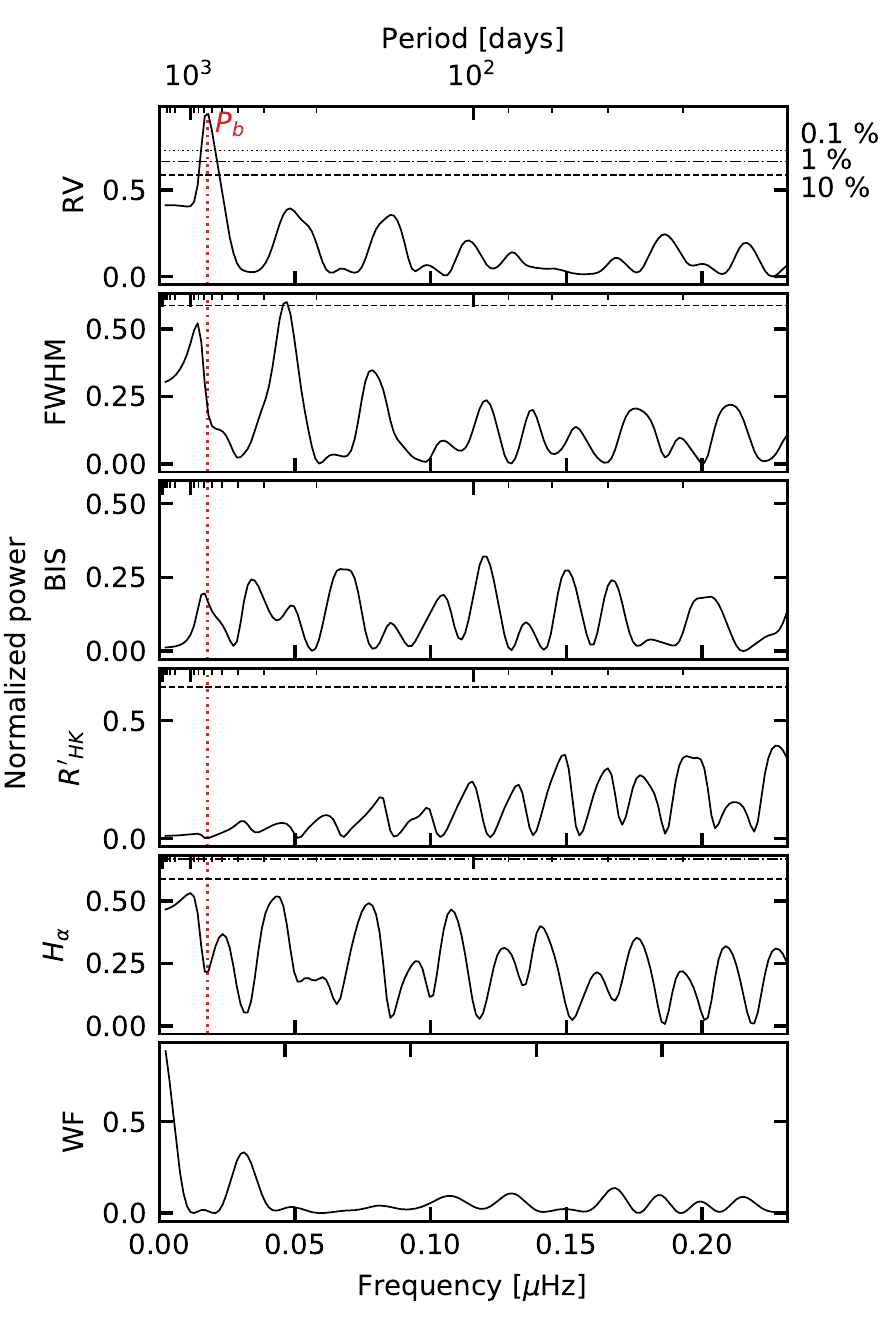}} %
    \caption{\figlabel{actinHD27969} \glsp\ of the \rv\ and the available activity indicators measured on HD27969 with the \sophie\ (\sophiep ) spectrograph.
    From top to bottom: \glsp\ of the \rv , \fwhm , \bis , and \logrhk\ time series and the window function of the data. The best-fit orbital period of the planet is marked by a vertical dashed red line. The black horizontal dotted, dash-dotted and dashed lines correspond to levels of 0.1, 1 and 10\,\% of \fap\ \citep{zechmeister2009}.}
\end{figure}

\begin{figure}[!htb]
    \centering
     \subfloat[HD80869 - \sophie\ and \sophiep]{\includegraphics[]{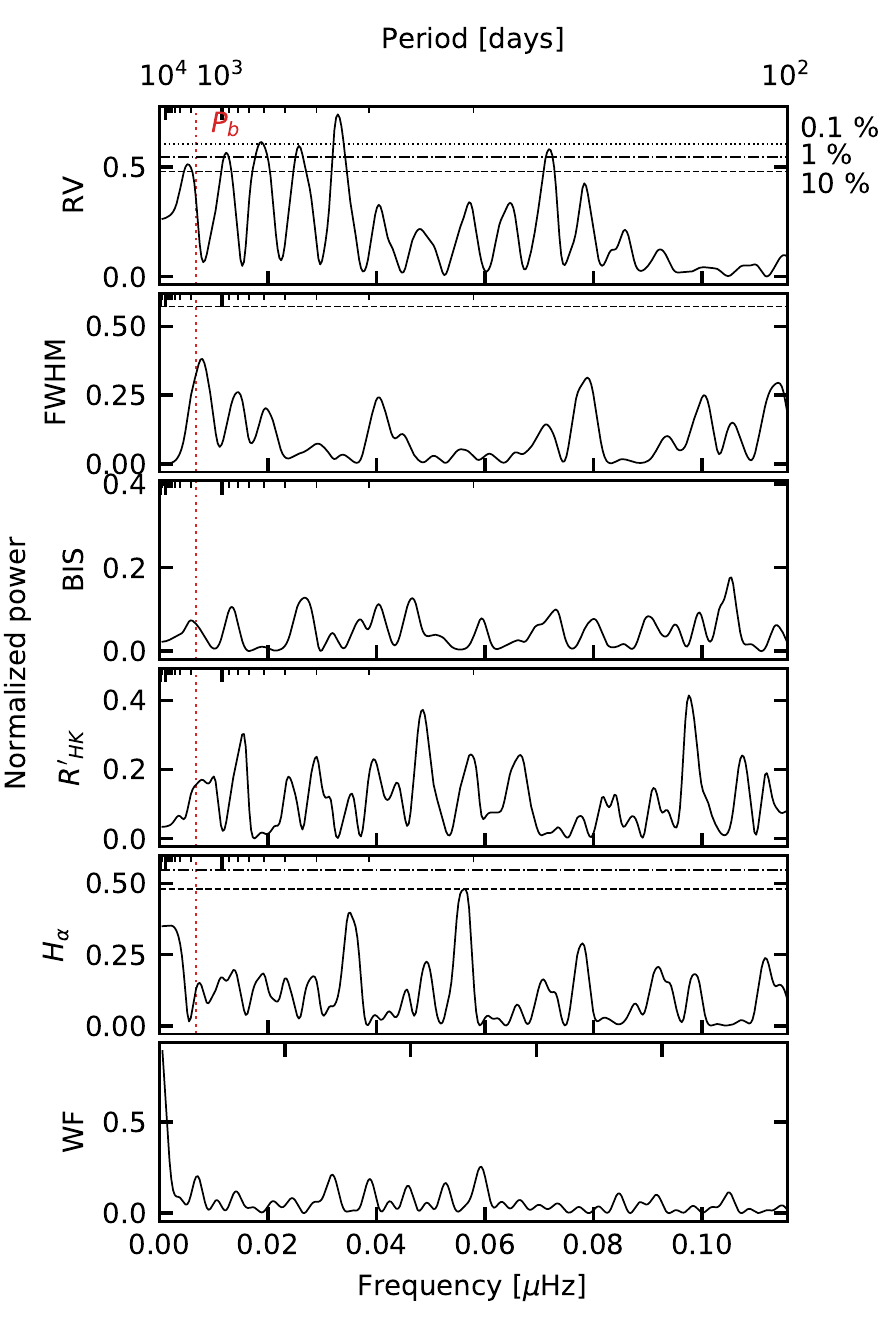}}
\hfil
    \subfloat[HD80869 - \elodie]{\includegraphics[]{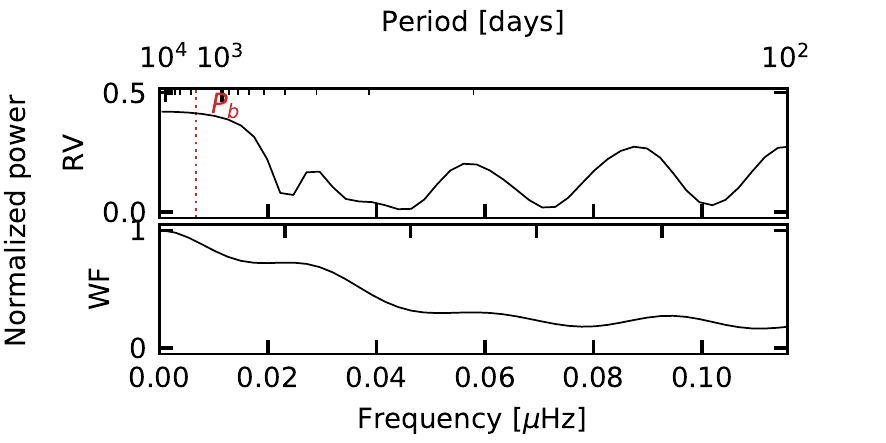}}
    \caption{\figlabel{actinHD80869} \glsp\ of the \rv\ and the available activity indicators measured on HD80869 with the \sophie\ (\sophie\ and \sophiep ) spectrograph (a) and the \elodie\ spectrograph (b). (a) From top to bottom: \glsp\ of the \rv , \fwhm , \bis , and \logrhk\ time series and the window function of the data. For the \glsp\ of the \rv\ data, the offset between \sophie\ and \sophiep\ is corrected prior to the computation of the \glsp\ using the value provided in \tab{sysparam}. This is not the case for the \fwhm , \bis,\ and \logrhk\ data since we do not have an estimate of the potential offset. (b) From top to bottom: \glsp\ of the \rv\ time series and the window function of the data. The rest of the format of this figure is the same as \fig{actinHD27969} (see the caption of that figure for more details).}
\end{figure}

\begin{figure}[!htb]
    \centering
    \subfloat[HD95544 - \sophiep]{\includegraphics[]{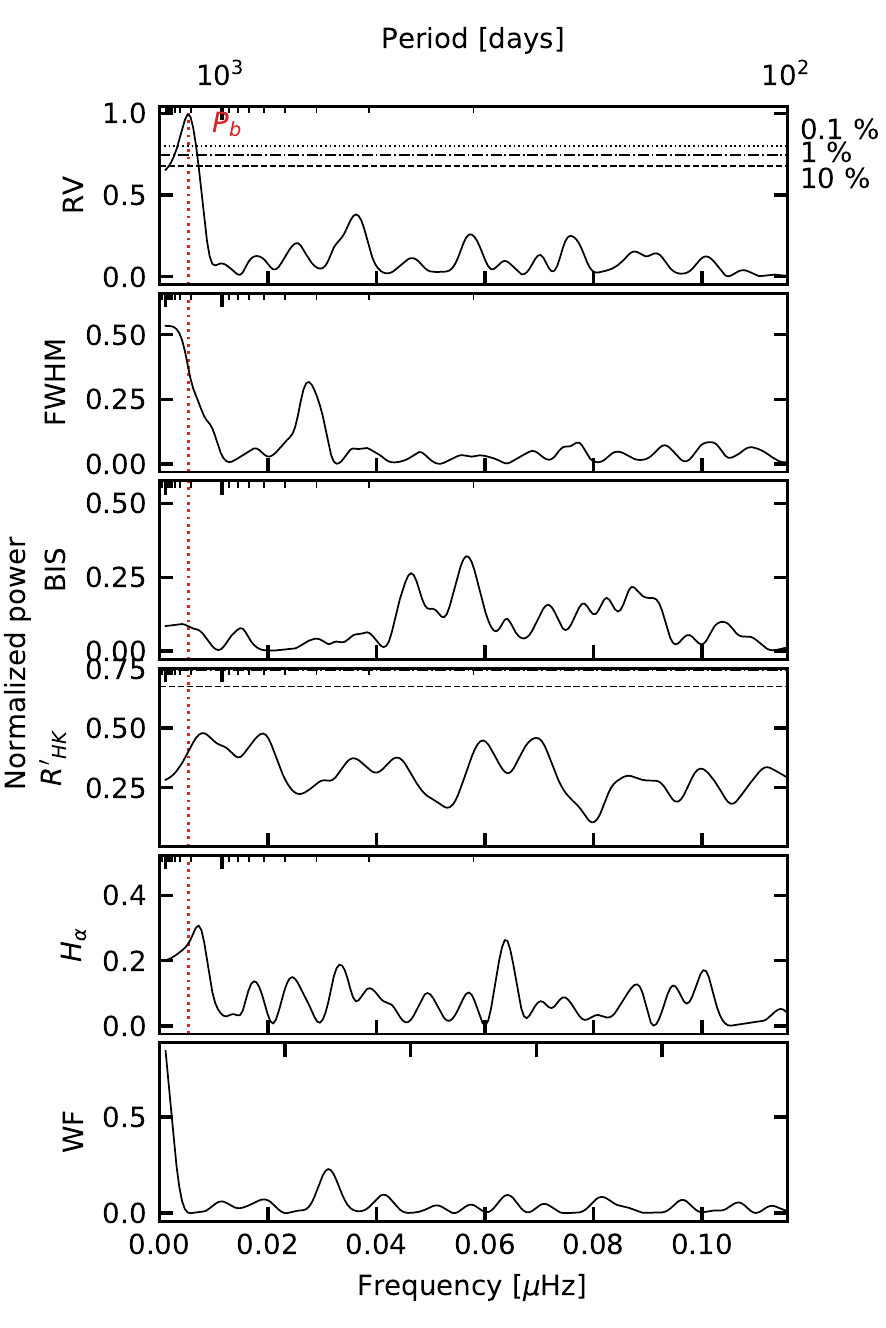}}
    \caption{\figlabel{actinHD95544} \glsp\ of the \rv\ and the available activity indicators measured on HD95544 with the \sophie\ (\sophiep ) spectrograph. From top to bottom: \glsp\ of the \rv , \fwhm , \bis , and \logrhk\ time series and the window function of the data. The rest of the format of this figure is the same as \fig{actinHD27969} (see the caption of that figure for more details).}
\end{figure}

\begin{figure}[!htb]
    \centering
    \subfloat[HD109286 - \sophiep]{\includegraphics[]{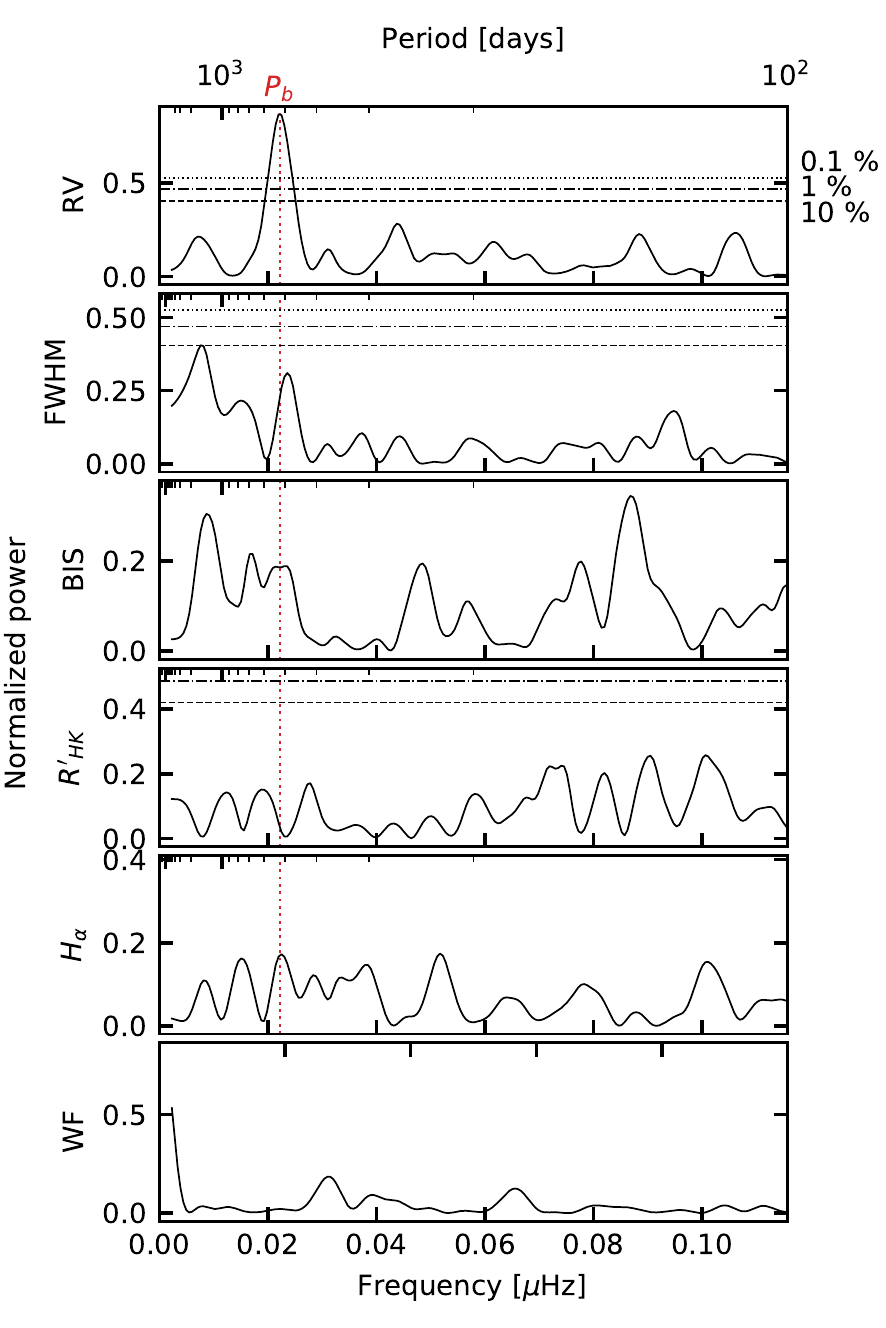}}
    \caption{\figlabel{actinHD109286}  \glsp\ of the \rv\ and the available activity indicators measured on HD109286 with the \sophie\ (\sophiep ) spectrograph. From top to bottom: \glsp\ of the \rv , \fwhm , \bis , and \logrhk\ time series and the window function of the data. The rest of the format of this figure is the same as \fig{actinHD27969} (see the caption of that figure for more details).}
\end{figure}

\begin{figure}[!htb]
    \centering
     \subfloat[HD115954 - \sophie\ and \sophiep]{\includegraphics[]{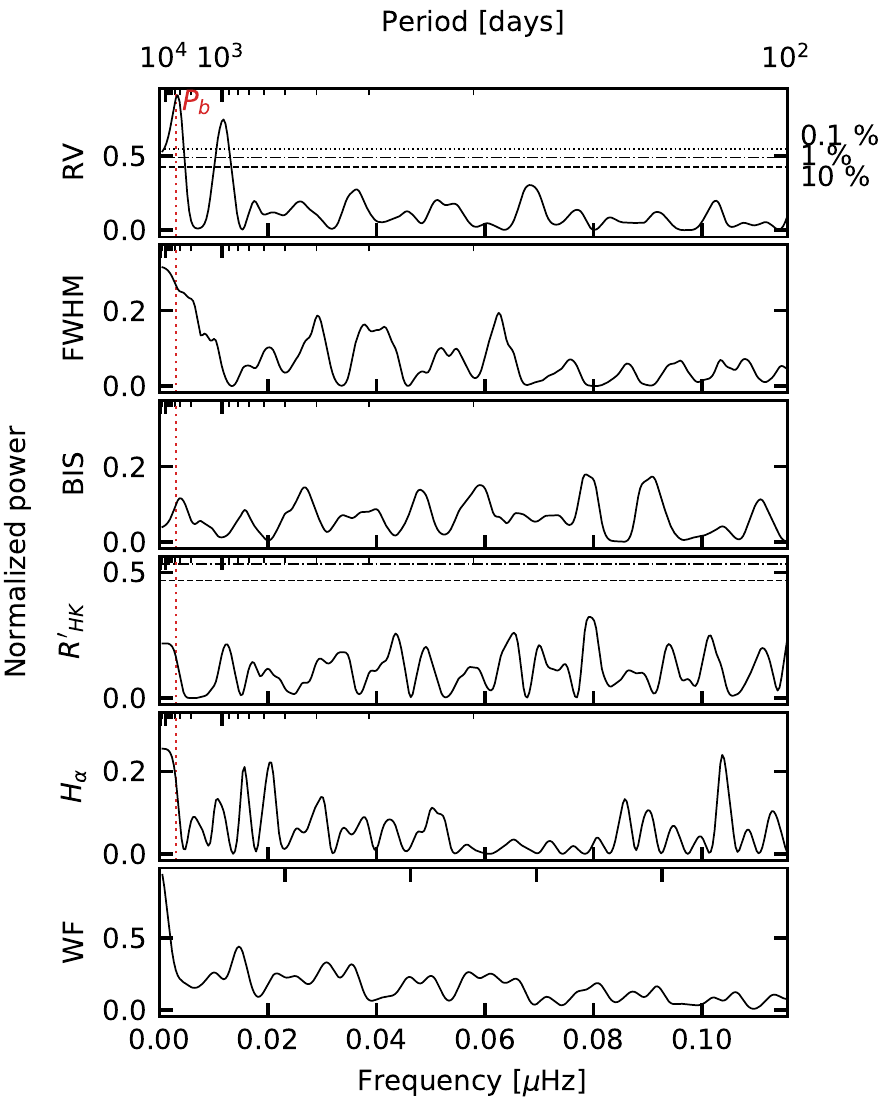}}
\hfil
    \caption{\figlabel{actinHD115954} \glsp\ of the \rv\ and the available activity indicators measured on HD115954 with the \sophie\ (\sophie\ and \sophiep ) spectrograph. From top to bottom: \glsp\ of the \rv , \fwhm , \bis , and \logrhk\ time series and the window function of the data. For the \glsp\ of the \rv\ data, the offset between \sophie\ and \sophiep\ is corrected prior to the computation of the \glsp\ using the value provided in \tab{sysparam}. Due to the limited number (four) of data points taken with \elodie , we do not present the \glsp\ as they do not contain much information. The rest of the format of this figure is the same as \fig{actinHD27969} (see the caption of that figure for more details).}
\end{figure}

\begin{figure}[!htb]
    \centering
     \subfloat[HD211403 - \sophie\ and \sophiep]{\includegraphics[]{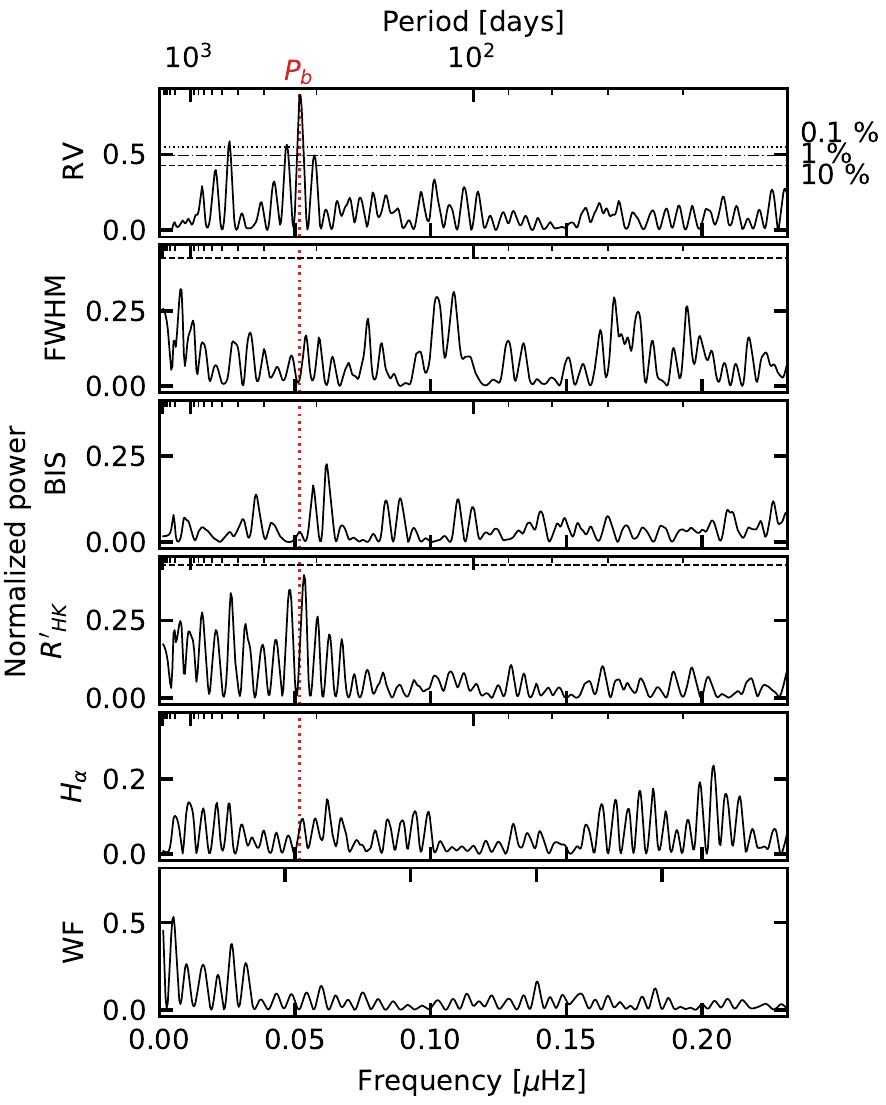}}
\hfil
    \subfloat[HD211403 - \elodie]{\includegraphics[]{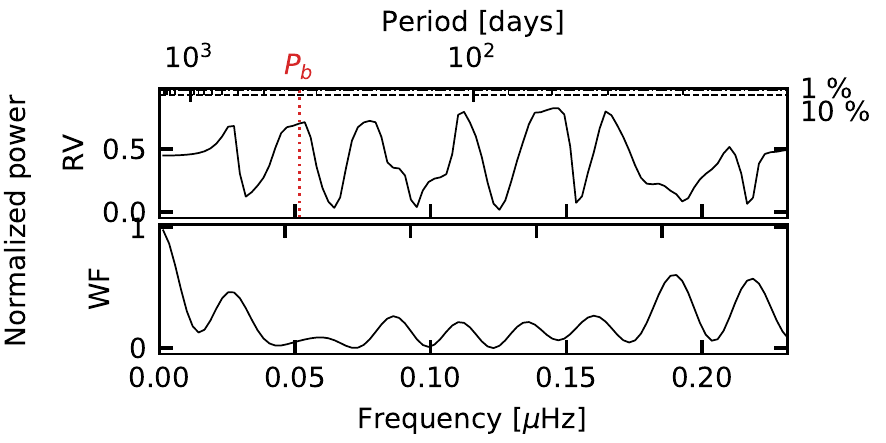}}
    \caption{\figlabel{actinHD211403} \glsp\ of the \rv\ and the available activity indicators measured on HD211403 with the \sophie\ (\sophie\ and \sophiep ) spectrograph (a) and the \elodie\ spectrograph (b). (a) From top to bottom: \glsp\ of the \rv , \fwhm , \bis , and \logrhk\ time series and the window function of the data. For the \glsp\ of the \rv\ data, the offset between \sophie\ and \sophiep\ is corrected prior to the computation of the \glsp\ using the value provided in \tab{sysparam}. This is not the case for the \fwhm , \bis,\ and \logrhk\ data since we do not have an estimate of the potential offset. (b) From top to bottom: \glsp\ of the \rv\ time series and the window function of the data. The rest of the format of this figure is the same as \fig{actinHD27969} (see the caption of that figure for more details).}
\end{figure}

\section{Radial velocities, \fwhm , \bis,\ and \logrhk}\applabel{rvs}

\tab{rvs27969} to \tab{rvs211403} provide the \rv , \fwhm , \bis , \logrhk\ and \halpha\  measurements obtained for our six stars. For \elodie\ data, we only derive \rv . We provide error bars for the \rv , \logrhk\ and \halpha\ measurements. Following \citet{Santerne2015}, the uncertainties on the \fwhm\ and \bis\ measurements can be obtained by multiplying the \rv\ uncertainties by a factor of 2.5.

%%%%%%%%%
% HD27969
%%%%%%%%%
\begin{table*}[!htb]
\tiny
\centering
\caption{\tablabel{rvs27969}\rv , \fwhm , \bis , \logrhk, and \halpha\ for star HD27969.}
\begin{tabular}{lccccc}
\hline
$\textrm{BJD}_{\textrm{UTC}}$ & RV    & FWHM  & BIS   & $log(\textrm{R'}_{\textrm{HK}})$ & H$_{\alpha}$ \\
- 2 400 000                   &       &       &       &                                  &              \\
days                          & \kms  & \kms  & \kms  &                                  &              \\
\hline\\[-5pt]

\multicolumn{6}{c}{Instrument: SOPHIE+} \\
\hline\\

57305.68205     & $47.2614 \pm 0.0031$ & 8.6844 & 0.0196  & $-5.117 \pm 0.038$ & $0.1048 \pm 0.0012$ \\
57316.60138     & $47.2787 \pm 0.0037$ & 8.6941 & 0.0012  & $-5.324 \pm 0.077$ & $0.1061 \pm 0.0015$ \\
57348.43339     & $47.2923 \pm 0.0037$ & 8.7090 & 0.0070  & $-5.375 \pm 0.087$ & $0.1030 \pm 0.0016$ \\
57364.47354     & $47.3100 \pm 0.0037$ & 8.7000 & 0.0016  & $-4.934 \pm 0.030$ & $0.1035 \pm 0.0016$ \\
\multicolumn{6}{l}{The full table is available in electronic form at the CDS ....} \\
\hline \hline
\end{tabular}
\tablefoot{\centering See text of \app{rvs} for the computation of the uncertainties of the \fwhm\ and \bis\ measurements.}
\end{table*}

%%%%%%%%%
% HD80869
%%%%%%%%%
%\pagebreak
\begin{table*}[!htb]
\tiny
\centering
\caption{\tablabel{rvs80869}\rv , \fwhm , \bis , \logrhk, and \halpha\ for star HD80869.}
\begin{tabular}{lccccc}
\hline
$\textrm{BJD}_{\textrm{UTC}}$ & RV    & FWHM  & BIS   & $log(\textrm{R'}_{\textrm{HK}})$ & H$_{\alpha}$ \\
- 2 400 000                   &       &       &       &                                  &              \\
days                          & \kms  & \kms  & \kms  &                                  &              \\
\hline\\[-5pt]

\multicolumn{6}{c}{Instrument: ELODIE} \\
\hline\\

53392.59240  & $-18.082 \pm 0.019$ & -- & -- & -- & -- \\
53724.66430  & $-18.245 \pm 0.014$ & -- & -- & -- & -- \\
53725.63520  & $-18.256 \pm 0.018$ & -- & -- & -- & -- \\
53749.66450  & $-18.239 \pm 0.019$ & -- & -- & -- & -- \\
\multicolumn{6}{l}{The full table is available in electronic form at the CDS ....} \\
\hline \hline
\end{tabular}
\tablefoot{\centering See text of \app{rvs} for the computation of the uncertainties of the \fwhm\ and \bis\ measurements. For the measurements made with \elodie , the \fwhm , \bis , \logrhk, and \halpha\ measurements are not available.}
\end{table*}

%%%%%%%%%
% HD95544
%%%%%%%%%
\begin{table*}[!htb]
\tiny
\centering
\caption{\tablabel{rvs95544}\rv , \fwhm , \bis , \logrhk, and \halpha\ for star HD95544.}
\begin{tabular}{lccccc}
\hline
$\textrm{BJD}_{\textrm{UTC}}$ & RV    & FWHM  & BIS   & $log(\textrm{R'}_{\textrm{HK}})$ & H$_{\alpha}$ \\
- 2 400 000                   &       &       &       & --                               & --           \\
days                          & \kms  & \kms  & \kms  & --                               & --           \\
\hline\\[-5pt]

\multicolumn{6}{c}{Instrument: SOPHIE+} \\
\hline\\

55906.72476 & $9.8201 \pm 0.0033$ &     7.9053 & -0.0052 & --               & $0.1103 \pm 0.0016$ \\
55929.67895 & $9.8113 \pm 0.0034$ &     7.9055 & -0.0104 & $-5.83 \pm 0.33$ & $0.1125 \pm 0.0016$ \\
56299.67406 & $9.7187 \pm 0.0033$ &     7.9145 & -0.0149 & $-5.41 \pm 0.11$ & $0.1130 \pm 0.0016$ \\
56370.56075 & $9.6956 \pm 0.0033$ &     7.9112 & -0.0056 & $-5.28 \pm 0.082$ & $0.1109 \pm 0.0016$ \\
\multicolumn{6}{l}{The full table is available in electronic form at the CDS ....} \\
\hline \hline
\end{tabular}
\tablefoot{\centering See text of \app{rvs} for the computation of the uncertainties of the \fwhm\ and \bis\ measurements.}
\end{table*}

%%%%%%%%%%
% HD109286
%%%%%%%%%%
\begin{table*}[!htb]
\tiny
\centering
\caption{\tablabel{rvs109286}\rv , \fwhm , \bis , \logrhk, and \halpha\ for star HD109286.}
\begin{tabular}{lccccc}
\hline
$\textrm{BJD}_{\textrm{UTC}}$ & RV    & FWHM  & BIS   & $log(\textrm{R'}_{\textrm{HK}})$ & H$_{\alpha}$ \\
- 2 400 000                   &       &       &       &                                  &              \\
days                          & \kms  & \kms  & \kms  &                                  &              \\
\hline\\[-5pt]

\multicolumn{6}{c}{Instrument: SOPHIE+} \\
\hline\\

56040.46588 & $-7.4998 \pm 0.0037$ & 8.0224 & 0.0081  & $-4.466 \pm 0.015$ & $0.1487 \pm 0.0017$ \\
56111.36864 & $-7.6357 \pm 0.0059$ & 8.1114 & -0.0039 & $-4.552 \pm 0.063$ & $0.1532 \pm 0.0015$ \\
56112.36722 & $-7.6393 \pm 0.0059$ & 8.1189 & 0.0048  & $-4.297 \pm 0.078$ & $0.1490 \pm 0.0014$ \\
56280.70866 & $-7.5019 \pm 0.0035$ & 7.9942 & -0.0116 & $-4.473 \pm 0.013$ & $0.1420 \pm 0.0017$ \\
\multicolumn{6}{l}{The full table is available in electronic form at the CDS ....} \\
\hline \hline
\end{tabular}
\tablefoot{\centering See text of \app{rvs} for the computation of the uncertainties of the \fwhm\ and \bis\ measurements.}
\end{table*}

%%%%%%%%%
% HD115954
%%%%%%%%%
% \pagebreak
\begin{table*}[!htb]
\tiny
\centering
\caption{\tablabel{rvs115954}\rv , \fwhm , \bis , \logrhk, and \halpha\ for star HD115954.}
\begin{tabular}{lccccc}
\hline
$\textrm{BJD}_{\textrm{UTC}}$ & RV    & FWHM  & BIS   & $log(\textrm{R'}_{\textrm{HK}})$ & H$_{\alpha}$ \\
- 2 400 000                   &       &       &       &                                  &              \\
days                          & \kms  & \kms  & \kms  &                                  &              \\
\hline\\[-5pt]

\multicolumn{6}{c}{Instrument: ELODIE} \\
\hline\\

53150.42490      & $-14.821     \pm 0.011$  & -- & -- & -- & -- \\
53152.39630      & $-14.838     \pm 0.009$  & -- & -- & -- & -- \\
53426.63970      & $-14.756     \pm 0.014$  & -- & -- & -- & -- \\
53428.63460      & $-14.788     \pm 0.028$  & -- & -- & -- & -- \\
\multicolumn{6}{l}{The full table is available in electronic form at the CDS ....} \\

\hline \hline
\end{tabular}
\tablefoot{\centering See text of \app{rvs} for the computation of the uncertainties of the \fwhm\ and \bis\ measurements. For the measurements made with \elodie , the \fwhm , \bis , \logrhk\ and \halpha\ measurements are not available.}
\end{table*}

%%%%%%%%%%
% HD211403
%%%%%%%%%%
\pagebreak
\begin{table*}[!htb]
\tiny
\centering
\caption{\tablabel{rvs211403}\rv , \fwhm , \bis , \logrhk, and \halpha\ for star HD211403.}
\begin{tabular}{lccccc}
\hline
$\textrm{BJD}_{\textrm{UTC}}$ & RV    & FWHM  & BIS   & $log(\textrm{R'}_{\textrm{HK}})$ & H$_{\alpha}$ \\
- 2 400 000                   &       &       &       &                                  &              \\
days                          & \kms  & \kms  & \kms  &                                  &              \\
\hline\\[-5pt]

\multicolumn{6}{c}{Instrument: ELODIE} \\
\hline\\
53216.58260      & $-9.514 \pm 0.036$  & -- & -- & -- & --\\
53218.55130      & $-9.440 \pm 0.030$  & -- & -- & -- & --\\
53333.32970      & $-9.292 \pm 0.045$  & -- & -- & -- & --\\
53585.55950      & $-9.109 \pm 0.074$  & -- & -- & -- & --\\
\multicolumn{6}{l}{The full table is available in electronic form at the CDS ....} \\

\hline \hline
\end{tabular}
\tablefoot{\centering See text of \app{rvs} for the computation of the uncertainties of the \fwhm\ and \bis\ measurements. For the measurements made with \elodie , the \fwhm , \bis , \logrhk, and \halpha\ measurements are not available.}
\end{table*}

% \section{Inferred stellar, planetary and instrumental parameters}

% \tab{sysparam} gathers the inferred stellar, planetary and instrumental parameters from the analyses described in \sect{plparam} to \sect{HD211403}.

% \input{table_syspar}

\end{appendix}

\clearpage
\onecolumn
\setcounter{table}{3}
\begin{landscape}
\begin{longtable}{lp{0.15\textwidth}p{0.15\textwidth}p{0.15\textwidth}p{0.15\textwidth}p{0.15\textwidth}p{0.15\textwidth}}
\caption{\tablabel{sysparam}System parameters from Bayesian MCMC analysis.}\\
%\hline
 & HD27969 & HD80869 & HD95544 & HD109286 & HD115954 & HD211403\\
\hline \\[-5pt]
\endfirsthead

\multicolumn{7}{c}{\textbf{\tablename\ \thetable{}} -- continued from previous page} \\[3pt]
 & HD27969 & HD80869 & HD95544 & HD109286 & HD115954 & HD211403 \\
\hline \\[-5pt]
\endhead

\hline
\multicolumn{7}{r}{Continued on next page\dots}\\
\multicolumn{7}{l}{See notes at the end of the table.}\\
\endfoot

\hline \hline

\endlastfoot

\multicolumn{7}{l}{\textit{Planetary parameters}} \\
\hline \\[-6pt]
%                                                                         & HD27969                              & HD80869                                & HD95544                               & HD109286                               & HD115954                              & HD211403 \\
$\mpl \sin{i_{\mathrm{p}}}$ [\mjup]                     & $4.80_{-0.23}^{+0.24}$         & $4.86_{-0.29}^{+0.65}$        & $6.84_{-0.31}^{+0.31}$                & $2.99_{-0.15}^{+0.15}$          & $8.29_{-0.58}^{+0.75}$        & $5.54_{-0.38}^{+0.39}$       \\[3pt]
$T_{\textrm{eq}}$ [K]                                           & $261_{-11}^{+11}$                         & $203.2_{-5.5}^{+6.8}$         & $156.5_{-5.5}^{+5.4}$                 & $259.4_{-5.5}^{+5.5}$                 & $144.9_{-13}^{+8.1}$                 & $3.80_{-13}^{+13}$       \\[3pt]
${P}^{\ \bullet}$\ [days]                                               & $654.5_{-5.8}^{+5.7}$           & $1711.7_{-9.6}^{+9.3}$                & $2172_{-21}^{+23}$              & $520.1_{-2.3}^{+2.3}$                 & $3700_{-390}^{+1500}$           & $223.8_{-0.41}^{+0.41}$       \\[3pt]
${t_{\textrm{ic}}}^{\bullet}$\ [BJD - 2\,455\,000]  & $2020.8_{-7.3}^{+7.3}$         & $1939_{-14}^{+12}$                    & $1191_{-15}^{+15}$                    & $1063.9_{-6.4}^{+6.2}$          & $701_{-53}^{+34}$             & $877.1_{-4.5}^{+5.2}$    \\[3pt]
$t_{\textrm{p}}$\ [BJD - 2\,455\,000]                   & $2042_{-16}^{+15}$                 & $1934_{-13}^{+13}$                    & $1700_{-160}^{+140}$                 & $1094.7_{-9.2}^{+8.6}$        & $1131_{-58}^{+49}$            & $929_{-54}^{+27}$    \\[3pt]
$a$ [AU]\footnotemark[1]                                & $1.552_{-0.033}^{+0.032}$     & $2.878_{-0.046}^{+0.045}$       & $3.386_{-0.077}^{+0.075}$     & $1.259_{-0.022}^{+0.021}$         & $5.00_{-0.36}^{+1.3}$                 & $0.768_{-0.013}^{+0.013}$       \\[3pt]
$e$                                                                     & $0.182_{-0.019}^{+0.019}$       & $0.862_{-0.018}^{+0.028}$     & $0.043_{-0.016}^{+0.017}$         & $0.338_{-0.035}^{+0.034}$     & $0.487_{-0.041}^{+0.095}$     & $0.084_{-0.044}^{+0.057}$      \\[3pt]
$\omega_*$ [$^\circ$]                                           & $106.6_{-7.7}^{+7.7}$         & $62.2_{-7.6}^{+4.2}$                  & $179_{-27}^{+0.23}$            & $133.1_{-7.4}^{+7.3}$                 & $185.6_{-8.8}^{+7.5}$                  & $190_{-57}^{+51}$       \\[3pt]
${e\cos \omega_*}^{\bullet}$                            & $-0.051_{-0.023}^{+0.023}$   & $0.396_{-0.053}^{+0.114}$   & $-0.040_{-0.017}^{+0.017}$    & $-0.228_{-0.041}^{+0.040}$         & $-0.483_{-0.084}^{+0.040}$    & $-0.058_{-0.066}^{+0.057}$       \\[3pt]
${e\sin \omega_*}^{\bullet}$                            & $0.172_{-0.020}^{+0.021}$         & $0.755_{-0.042}^{+0.035}$     & $0.001_{-0.017}^{+0.017}$     & $0.244_{-0.037}^{+0.037}$       & $-0.047_{-0.081}^{+0.072}$    & $-0.011_{-0.052}^{+0.054}$       \\[3pt]
$a / R_*$                                                               & $263_{-20}^{+23}$                       & $583_{-24}^{+0.26}$           & $669_{-45}^{+50}$                       & $248.5_{-9.9}^{+11}$          & $903_{-98}^{+220}$              & $136.9_{-8.8}^{+9.9}$       \\[3pt]
${K}^{\ \bullet}$ [\ms]                                         & $103.6_{-2.6}^{+2.8}$         & $151.9_{-9.5}^{+40}$                  & $101.5_{-1.6}^{+1.6}$                  & $81.6_{-2.7}^{+2.8}$          & $110.7_{-6.1}^{+6.9}$                  & $165.5_{-10.0}^{+9.8}$       \\[3pt]
${\textrm{RV slope}}^{\bullet}$ [\ms.$y^{-1}$]                  &                                                &                                                &                                                       & $3.7_{-1.3}^{+1.3}$             & $0.9_{-2.0}^{+1.8}$                   &        \\[3pt]
$F_{i}$ [$F_{i, \oplus}$]                                       & $0.76_{-0.12}^{+0.13}$         & $0.142_{-0.012}^{+0.013}$     & $0.100_{-0.013}^{+0.015}$     & $0.709_{-0.058}^{+0.061}$       & $0.064_{-0.023}^{+0.017}$     & $3.44_{-0.46}^{+0.50}$       \\[3pt]

\\[-3pt]
\multicolumn{7}{l}{\textit{Stellar parameters}} \\
\hline \\[-5pt]
%                                                                               & HD27969                 & HD80869                       & HD95544                        & HD109286                      & HD115954                      & HD211403 \\
\textsc{ra}$^{\textsc{gaia-crf2}}$  [hh:mm:ss.ssss]     & 4:26:49.6407           & 9:23:09.9103                  & 11:05:17.5628                 & 12:33:35.0020                   & 13:19:56.4887                 & 22:15:15.7963     \\[3pt]
\textsc{dec}$^{\textsc{gaia-crf2}}$ [dd:mm:ss.ss]       & 42:54:31.39            & 33:54:17.39                   & 81:02:20.35                   & 07:16:51.26                     & 38:22:08.90                   & 58:16:32.02       \\[3pt]
Sp. Type                                                                        & G0V                     & G1V                           & G3V                                    & G4V                           & G0V                            & F7V               \\[3pt]
V mag                                                                           & 7.65                    & 8.45                          & 8.39                           & 8.77                          & 8.34                           & 8.50              \\[3pt]
$M_*$ [M$_{\sun}$]                                                      & $1.16 \pm 0.07$         & $1.08 \pm 0.05$               & $1.09 \pm 0.07$                 & $0.98 \pm 0.05$               & $1.18 \pm 0.06$               & $1.20 \pm 0.06$   \\[3pt]
$R_*$ [R$_{\sun}$]                                                      & $1.27 \pm 0.10$         & $1.062 \pm 0.042$             & $1.088 \pm 0.072$             & $1.089 \pm 0.041$               & $1.213 \pm 0.083$             & $1.206 \pm 0.080$ \\[3pt]
\teff\ [K]                                                                      & 5966 $\pm$ 21           & 5837 $\pm$ 15                 & 5722 $\pm$ 15                  & 5694 $\pm$ 23                 & 5957 $\pm$ 26                 & 6273 $\pm$ 44    \\[3pt]
\logg\ (dex)                                                                    & 4.12 $\pm$ 0.04 & 4.18 $\pm$ 0.03               & 4.07 $\pm$ 0.03               & 4.44 $\pm$ 0.04                 & 4.15 $\pm$ 0.04               & 4.30 $\pm$ 0.15  \\[3pt]
{[Fe/H]} [dex]                                                                  & 0.18 $\pm$ 0.02         & 0.17 $\pm$ 0.01               & 0.11 $\pm$ 0.01                 & 0.05 $\pm$ 0.02               & 0.34 $\pm$ 0.02               & 0.18 $\pm$ 0.04 \\[3pt]
\vsini\ [\kms]                                                                  & $4.1 \pm 1$                             & $3.7 \pm 1$                                   & $3.3 \pm 1$                                     & $3.8 \pm 1$                                    & $5.3 \pm 1$                                   & 18.4 $\pm$ 0.3  \\[3pt]
vturb [\kms]                                                        & 1.17 $\pm$ 0.02      & 1.02 $\pm$ 0.02               & 1.02 $\pm$ 0.02               & 1.01 $\pm$ 0.03                 & 1.15 $\pm$ 0.03               &                 \\[3pt]
\logrhk\ [dex]                                                       & -5.3              & -5.1                             & -5.2                          & -4.45                           & -5.1                          & -4.6               \\[3pt]

%                                                                                & HD27969                                       & HD80869                                        & HD95544                                       & HD109286                                        & HD115954                                & %HD211403 \\
${v_{0,\textrm{SOPHIE+}}}^{\bullet}$ [\kms]                     & $47.2848_{-0.0017}^{+0.0017}$ & $-18.1373_{-0.0026}^{+0.0034}$  & $9.7509_{-0.0012}^{+0.0012}$  & $-7.5227_{-0.0044}^{+0.0045}$ & $-14.7444_{-0.0095}^{+0.0174}$  & $-9.1262_{-0.0074}^{+0.0078}$  \\[3pt]

\\[-5pt]
\multicolumn{7}{l}{\textit{Parameters of instruments}} \\
\hline \\[-5pt]
%                                                                                                       & HD27969                         & HD80869                       & HD95544                         & HD109286                      & HD115954                       & HD211403 \\
${\Delta\textrm{RV}_{\textrm{SOPHIE/SOPHIE+}}}^{\bullet}$ [\ms]         &                                         & $12.2_{-7.5}^{+6.5}$      &                                    &                                       & $1_{-13}^{+12}$                 & $-9_{-17}^{+17}$       \\[3pt]
${\Delta\textrm{RV}_{\textrm{ELODIE/SOPHIE+}}}^{\bullet}$ [\ms]         &                                         & $-37.4_{-8.2}^{+8.3}$     &                                    &                                       & $-79_{-48}^{+56}$               & $-219_{-29}^{+30}$ \\[3pt]
$\sigma_{\textrm{SOPHIE+}}^{\bullet}$ [\ms]                                     & $5.1_{-1.3}^{+1.7}$     & $6.46_{-0.90}^{+1.09}$  & $3.9_{-1.1}^{+1.3}$ & $13.1_{-1.5}^{+1.9}$    & $6.7_{-1.1}^{+1.4}$        & $35.2_{-5.6}^{+6.8}$               \\[3pt]
$\sigma_{\textrm{SOPHIE}}^{\bullet}$ [\ms]                                      &                                         & $< 11$                                &                                         &                                       & $< 6.8$                         & $40_{-10}^{+15}$        \\[3pt]
$\sigma_{\textrm{ELODIE}}^{\bullet}$ [\ms]                                      &                                         & $29.4_{-5.6}^{+7.1}$      &                                    &                                       & $29_{-16}^{+18}$        & $61_{-29}^{+36}$        \\[3pt]
\end{longtable}
\tablefoot{\\
-- $^{\textsc{gaia-crf2}}$ indicates that the coordinates are provided in the equatorial system with the Gaia Celestial Reference Frame \citep[Gaia-CRF2][]{Gaia2018}.\\
-- Spectral Types are estimated from \teff\ using the Table 5 of \citet{Pecaut2013}.\\
-- For $M_*$ and $R_*$ only statistical errors are included.\\
-- The best-fit values are computed as the median of the cleaned MCMC chains.\\
-- The errors bars correspond to 68\%\,confidence intervals.\\
-- The \vsini\ values were obtained from the \textsc{ccf} using the \textsc{sophie} pipeline \citep[][]{boisse2010}, except for the fast rotator HD211403 for which the \textsc{sme} software was used \citep[][]{valenti1996}. However, we not that for this star \citep[][]{boisse2010} provides an estimate of $17.9 \pm 1\,\kms$ in good agreement with the \textsc{sme} value.\\
-- The micro-turbulence velocity ($v$turb) is estimated as part of the spectroscopic stellar parameter derivation (\sect{spectro}). We refer the reader to \citet[][]{sousa2014} for more details.\\
-- The reference time for the \rv\ slope model is the time of the first data point available in the data set.\\
-- $^{\bullet}$ indicates that the parameter is a main or jumping parameter for the \textsc{mcmc} explorations performed in \sect{plparam}.\\
-- \footnotemark[1] $M_p$ is inferred assuming an inclination of $\arcsin(\sqrt{\pi / 4})$ which is the average inclination assuming that the inclination are uniformly distributed in space \citep{lovis2010}. $a$ is computed from $a / R_*$.\\
}
\end{landscape}

\end{document}